\documentclass[preprint,5p,times,twocolumn,number]{elsarticle}
\usepackage{hyperref}
\usepackage[english]{babel}
\usepackage[usenames, dvipsnames]{color}
\usepackage{graphicx}
\usepackage{tikz,lipsum,lmodern}
\usepackage[most]{tcolorbox}
\usepackage{array,tabularx,enumitem,booktabs}
\newcolumntype{Y}{>{\raggedright\arraybackslash}X}
\setlist{nosep}
\usepackage [dvipsnames]{xcolor}

\usepackage{supertabular}
\usepackage{longtable}
\usepackage{multirow}
\usepackage{array}
\usepackage{url}
\usepackage{subcaption}
\usepackage{subcaption}
\usepackage[utf8]{inputenc}
\usepackage[export]{adjustbox}
\usepackage{wrapfig}
\usepackage{multicol}
\usepackage{float}
\usepackage{todonotes}
\usepackage[T1]{fontenc}
\usepackage{numcompress}
\usepackage{comment}

\usepackage{makecell}

\journal{Journal of Systems and Software}

\makeatletter
\def\@author#1{\g@addto@macro\elsauthors{\normalsize%
    \def\baselinestretch{1}%
    \upshape\authorsep#1\unskip\textsuperscript{%
      \ifx\@fnmark\@empty\else\unskip\sep\@fnmark\let\sep=,\fi
      \ifx\@corref\@empty\else\unskip\sep\@corref\let\sep=,\fi
      }%
    \def\authorsep{\unskip,\space}%
    \global\let\@fnmark\@empty
    \global\let\@corref\@empty  
    \global\let\sep\@empty}%
    \@eadauthor={#1}
}
\makeatother

\begin{document}

\begin{frontmatter}

\title{Software-Defined Vehicle Ecosystems in Transformation - A Systematic Literature Review}


\author{Heidi Hietala\corref{cor1}}
\ead{HeidiHietala@oulu.fi}
\cortext[cor1]{Corresponding author}

\author{Nirnaya Tripathi, Prabhash Rathnayake, Yueqiang Xu,  Tero Päivärinta, and Ella Peltonen} 

\address{M3S Research Group, University of Oulu, Finland}

\begin{abstract}
\textbf{Context:} The automotive industry is shifting from hardware-centric development toward software-defined vehicles (SDVs), where software drives functionality, value creation, and competitive differentiation. Growing software complexity renders firm-centric and proprietary software development models insufficient, prompting a shift toward ecosystem collaboration among OEMs, suppliers, and software firms. Yet, how these SDV ecosystems emerge and operate in response to software-driven development remains insufficiently understood.

\textbf{Objectives:} This study enhances our understanding of SDV ecosystems, outlines their collaborative structures, identifies stakeholders, their roles and authority, and highlights associated challenges and opportunities.

\textbf{Methods:} This systematic literature review collected 351 studies from six databases and selected 25 for detailed analysis.

\textbf{Results:} This study identifies six levels of collaboration involving twelve stakeholder groups shaping SDV ecosystem transformation. These collaborations are influenced by five dimensions of authority. SDV ecosystems face six core software development challenges alongside six organisational, six industry and market, and four regulatory, legal, and ethical challenges. The literature also highlights five key software development opportunities complemented by six organisational, four industry and market, and two public value and ethical opportunities. 

\textbf{Conclusion:} SDV ecosystem research is primarily technical, concentrating on architectures and standardisation, while lacking studies on governance and collaborative software business models that reflect regional characteristics and power dynamics. We reposition SDVs as multi-level socio-technical ecosystems where software functions as the core structuring principle but does not alone determine ecosystem success. We develop a multi-level SDV ecosystem model, integrating stakeholders, collaborative structures, and governance across ecosystem levels, and outline directions for future research and practice.

\end{abstract}



\end{frontmatter}



\section{Introduction}
\label{sec1}

The automotive industry is undergoing a significant transformation from traditional hardware-focused engineering and supply-chain-based business models to new business-to-consumer approaches that emphasise customer engagement and supplier partnerships through data \cite{Paulweber2025Multi-PartnerEcosystem, Llopis-Albert2021ImpactIndustry}. This paradigm shift redefines vehicles as digital platforms on wheels, where the sale of the vehicle marks just the beginning of ongoing applications and connections for continuous revenues \cite{Lu2023VehicleChallenges}. Central to this transformation is the software-defined vehicle (SDV) concept, which abstracts software from hardware, enabling continuous adaptation and improvement. Unlike traditional vehicles, which focus on mechanical components and embedded software, SDVs utilise Service-Oriented Architecture (SoA) to deploy various hardware units based on their capabilities, service needs, and current network status.

An SDV can be defined as \textit{"an automotive vehicle whose core value proposition (e.g., autonomous driving) is predominantly enabled, controlled, and continually enhanced by software rather than mechanical components"} \cite{Otto2025TowardsAvenues}. These vehicles are powered by advances in key technologies, such as cloud and edge computing \cite{Paulweber2025Multi-PartnerEcosystem}, centralised electrical and/or electronic (E/E) architectures \cite{Bandur2021MakingArchitectures}, high-performance computers (HPCs)\cite{Windpassinger2022OnVehicle}, and over-the-air (OTA) software updates \cite{Malik2022Over-the-AirEnvironment}. These advances enable implementing various functional and service applications on top of the basic software \cite{Liu2022ImpactVehicles}. The applications range from hardware control and security features to various service and infotainment functionalities \cite{Liu2022ImpactVehicles}. Like smartphones, an SDV can be personalised and updated with the latest software and firmware throughout the vehicle's lifecycle, integrating components from various suppliers \cite{Lu2023VehicleChallenges}.

Software development is the core enabler of SDVs, shifting vehicles from static mechanical products to continuously evolving digital platforms where software defines functionality, performance, safety, and user experience. This shift requires large, high-quality software systems that evolve over the vehicle lifecycle through practices such as continuous integration, continuous delivery, AI/ML services and OTA updates, supported by real-world data from deployed vehicles \cite{HillierIonescu2025CloudSDV, HillierIonescu2025SDVSuccess}. However, the shift substantially increases software complexity, lifecycle duration, and interdependencies across hardware, cloud infrastructure, data platforms, and third-party services—exceeding the capacity of firm-centric development models.

\begin{figure}
    \centering
\includegraphics[width=.8\linewidth]{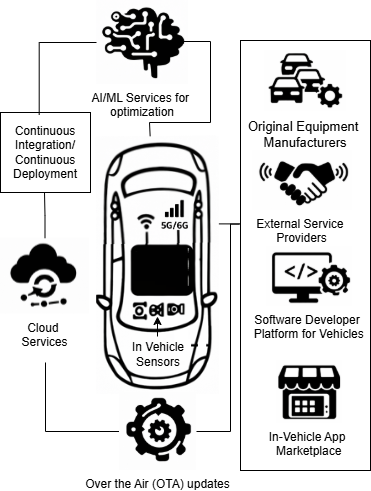}
    \caption{Overview of a software-defined vehicle and its ecosystem aspects}
    \label{fig:SDV}
\end{figure}

The technological advances that comprise the SDV, with complex webs of dependencies, transform the entire value chain \cite{liebetrau2022architecture}. Traditionally, OEMs primarily interact with specific tier companies, but will now evolve into more interconnected networks with direct collaborations across multiple tiers \cite{liebetrau2022architecture}. Subsequently, SDVs foster the emergence of new ecosystems by combining open-source and standardised technologies with novel vehicle architecture and technologies \cite{Paulweber2025Multi-PartnerEcosystem, khamis2025rethinking}. These ecosystems allow stakeholders to contribute data and services that enhance passenger comfort, safety, and overall user experience \cite{Cheng2025Anti-Kinetosis:Vehicles}. Moreover, by connecting hardware and software suppliers and third-party developers, the automotive industry can access a broader application ecosystem \cite{Liu2022ImpactVehicles}. Consequently, software development for SDVs increasingly depends on ecosystem-level collaboration, in which architectures and development practices are shaped by shared standards, partnerships, and governance mechanisms that support interoperability and continuous delivery. Figure \ref{fig:SDV} illustrates this overview of the SDV and ecosystem actors.

However, the shift to SDV architectures presents socio-technical challenges and requires new organisational structures \cite{Panchal2023DesignOpportunities} and novel business models \cite{Liu2022ImpactVehicles}. New OEMs, such as Lucid, Rivian, and Nio, advance rapidly with their software-first approach, while traditional manufacturers, like Volkswagen, Mercedes, General Motors, and Ford, are facing challenges in transforming their hardware-centric models, largely due to legacy systems and practices \cite{khamis2025rethinking}. Additionally, balancing the fast-paced software culture, like that in Silicon Valley, with the traditional automotive focus on rigorous testing to ensure safety and reliability for frequently updated software is critical, as both have limitations \cite{Panchal2023DesignOpportunities}. Thus, the automotive industry must work to transition from proprietary software and siloed supply chains to a more connected, collaborative ecosystem involving multiple partners with differing interests \cite{Peltonen2024Software-DefinedReport}. While similar industry transformations are taking place elsewhere, e.g., digital twins \cite{tripathi2024stakeholders}, to realise these visions, network-based collaboration in ecosystems is essential. 

Despite increasing research interest in SDVs, our understanding of their ecosystems remains limited \cite{Otto2025TowardsAvenues}. While parallels are made to the transformation of the mobile phone industry \cite{becker2022operating, Lu2023VehicleChallenges}, the safety-critical requirements and long life cycles of SDVs demand a unique approach to their development and regulation. Understanding the ecosystems' structures and dynamics is critical for supporting the industry's transition toward software-defined mobility. 

Accordingly, we set five research questions (Table \ref{RQ}) with \textbf{the main objective of shedding light on current knowledge of SDV ecosystems.} More specifically, we aim to explore the collaborative structures of these ecosystems, identify the stakeholders involved, their roles and authority, as well as the emerging challenges and opportunities for the automotive software industry and beyond. 

\begin{table*}[h!]
\caption{Research Questions}
\fontsize{9}{10}\selectfont
\label{RQ}
\begin{tabular}{|p{5.5cm}|p{11.9cm}|}

\hline
\textbf{Research Questions}  & \textbf{Rationale}                                      \\ \hline

        \textbf{RQ1 - }What is currently known about SDV ecosystem research? & Given the novelty of the SDV concept, an overview of its ecosystems is needed to synthesise existing knowledge and identify dominant themes and gaps. \\ \hline
        \textbf{RQ2 - }How are collaborative structures organised in SDV ecosystems? & SDVs require collaboration and coordination across multiple stakeholders and domains, making it necessary to understand how collaboration is structured beyond traditional automotive supply chains. \\ \hline
        \textbf{RQ3 - }Who are the stakeholders in SDV ecosystems, and what are their roles and authorities? & SDV ecosystems consist of a diverse and expanding set of stakeholders, making it essential to understand their roles, responsibilities, and decision-making authority, especially as these aspects continue to evolve. \\ \hline
        \textbf{RQ4 - }What are the main challenges faced by SDV ecosystems? & The transition to SDVs introduces new intertwined challenges that must be identified to understand barriers to ecosystem collaboration and value creation. \\ \hline
        \textbf{RQ5 - }What are the main opportunities presented by SDV ecosystems? & The transition to SDVs creates new opportunities that must be identified to understand how ecosystem collaboration and value creation can be enabled. \\ \hline

    \end{tabular}
\end{table*}

To achieve these objectives, we conducted a systematic literature review (SLR) \cite{Kitchenham2004ProceduresReviews, Kitchenham2010SystematicStudy} focusing on studies that explore different aspects of collaboration, competition, and value co-creation within automotive software in relation to SDV ecosystems. We drew 351 studies from six databases, and selected 25 as the primary studies for our analysis.

The SLR findings outline six levels of collaboration involving twelve stakeholder groups that influence the transformation of the SDV ecosystem. These collaborative structures are shaped by five dimensions of authority: control centralisation, power redistribution, platformisation, and the trade-offs between control and openness, as well as the impacts of regulatory, regional, and political factors. The transformation involves six primary software grand challenges. Additionally, six organisational, six industry and market, and four regulatory, legal, and ethical challenges were identified. In a similar vein, SDV ecosystems offer five key software opportunities, complemented by six organisational, four industry and market, and two public value and ethical opportunities. Collectively, these insights consolidate fragmented research on the SDV ecosystem into a cohesive perspective on its transformation.

Our SLR reveals that studies on the SDV ecosystem are largely focused on technical aspects, emphasising architectures, standardisation, and interoperability. However, SDV ecosystems also encompass multi-level collaboration among various stakeholders, power dynamics, and evolving business models for value creation. We contend that SDVs should be viewed as complex socio-technical ecosystems, where software serves as the core structuring principle, but is not the sole factor influencing SDV ecosystems' success. Therefore, this study proposes a multi-level SDV ecosystem model that incorporates stakeholders, collaborative structures, and governance across different levels of the ecosystem. This study underscores the necessity for additional empirical research into governance, software business models, and regional variation, while also outlining practical implications for advancing SDV ecosystems.
\section{Background and Related Work}

The concept of SDVs has gained significant research interest, with literature reviews exploring their implications and applications (cf. \cite{mate2025software, Otto2025TowardsAvenues, khamis2025rethinking}). However, there is still much to investigate regarding their ecosystem \cite{Otto2025TowardsAvenues, khamis2025rethinking}. This section will outline the key technological foundations, software aspects, collaborative business models, and the overall SDV ecosystem.

\subsection{Technological Foundations of SDVs}

An SDV, as a technical term, can be seen as a by-product of the increased amount of software required to run a modern vehicle, from infotainment to Advanced Driver Assistance Systems (ADAS) technologies, such as lane-following assistance, automated braking, and adaptive cruise control. Sensors and cameras have become integral for an enjoyable and safer driving experience, all of which require extensive amounts of software running in real-time and managing electronic control units (ECUs) corresponding to sensing and actuator roles on vehicles. 

ECUs are highly embedded by design, integrating low-abstraction software and hardware components from Tier 1 companies. Their integration, communication buses, and positioning in the chassis are often specific to each OEM. \cite{coppola2016connected}. In general, the orchestration or organisation of ECUs compromises the vehicle architecture, which is nowadays distributed across physical, functional, or domain-specific zones connected through gateways \cite{khamis2025rethinking}. Thus, it would be a mistake to call a vehicle "a computer" instead of a distributed network of highly specialised, embedded computers, each of which performs its computations in a safety-critical, cyber-physical environment. 

Vehicle communication relies on the embedded and distributed nature of ECUs. Most vehicles implement a number of buses, most importantly, Controller Area Network (CAN), responsible for drivetrain and safety systems, and Local Interconnect Networks (LIN) for locally interconnected ECUs. The CAN-based architectures tend to be cheaper and easier to structure, as every ECU efficiently (or through a gateway) listens to the same communication pipe. As CAN bus are seen as a legacy system with limited data rate and limitations in security, suggestions have been made for automotive-grade Ethernet (e.g., IEEE 802.3bw / 802.3bp / 802.3ch), an adaptation of the traditional networking model designed for computer clusters, to serve as a communication base for in-vehicular communication \cite{douss2023state}. The change in networking backbone would also provide access for more complex architectures, as the number of ECUs is indeed increasing. However, buses are deeply embedded in the industry, inexpensive, and remain effective for low-level and local communication tasks.

\subsection{Software Aspects of SDVs}

\begin{figure*}
    \centering
    \includegraphics[width=0.9\linewidth]{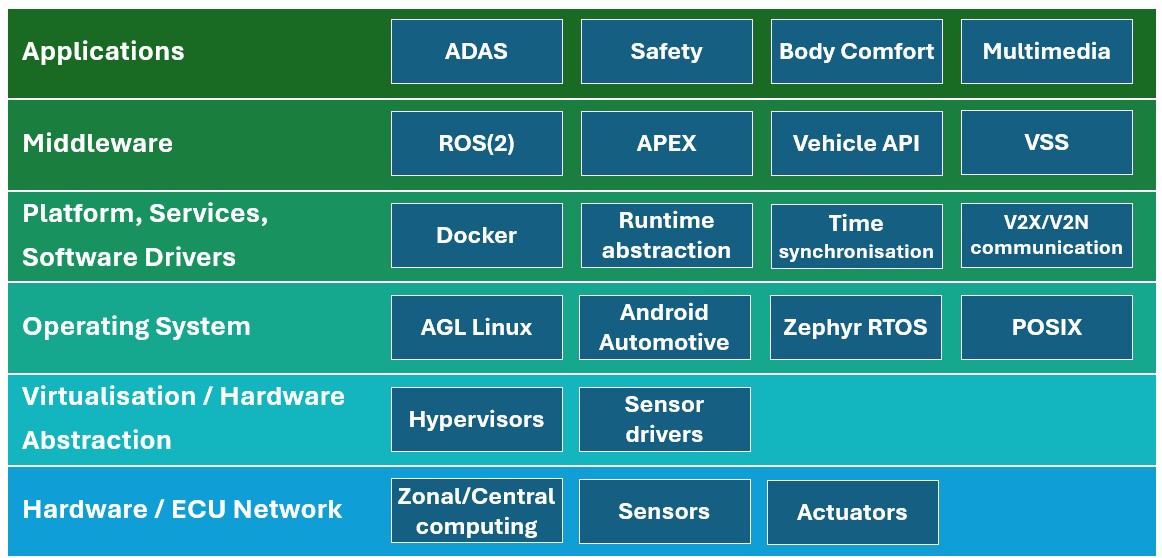}
    \caption{Example of a generic SDV architecture \cite{Paulweber2025Multi-PartnerEcosystem, liu2022impact}.}
    \label{fig:architecture}
\end{figure*}

The long lifespan of vehicles over decades, requires careful attention to software engineering, including continuous testing, validation, and verification practices. Key challenges in vehicular software development include, but are not limited to, integrating software and various hardware components, ensuring compatibility and code reusability across complex ECU architectures, and addressing the safety-critical nature of vehicle software, which demands reliability and security \cite{moukahal2020vehicle}. Additionally, technological fragmentation into manufacturer-specific silos increases software development risks and complicates the entry of new stakeholders into the market, necessitating software support across multiple platforms. 

There is no unified understanding of the SDV's software architecture, but a general framework can be drawn from a hardware-to-application tier architecture, as shown in Figure \ref{fig:architecture}. The abstraction level increases from bottom up: from hardware and the ECU network connected by different buses through hardware virtualisation, operating systems, various microservice-related software drivers, and programmable middleware, to the application layer. Currently, these architectures are typically manufacturer-specific. However, some geopolitical alliances and government initiatives for shared SDV platforms exist at least in Japan and the European Union (EU) \cite{Peltonen2024Software-DefinedReport}. A common trend is the shift towards microservices, containerisation, and service-oriented architectures \cite{kugele2017service,vetter2020development}.

Unlike most software, vehicular management and update cycles are still closely tied to mechanical maintenance through predefined syncs. From the modern software engineering perspective, such pre-planned maintenance is outdated, leading to a focus on OTA updates for SDVs. More frequent OTAs could introduce new features, applications, and capabilities, thereby increasing vehicle market value without mechanical updates \cite{wang2025exploring}. However, OTA technology is notorious for security risks and potential attacks \cite{mahmood2022systematic}, underscoring the urgent need for rigorous safety and security verification and validation throughout the software development cycle. Management of the full vehicular software lifecycle, supported by OTA updates, remains a key challenge for SDV development \cite{halder2020secure}. 

\subsection{Collaborative Business Models and SDV Ecosystems}

The automotive industry's transformation through SDVs is reshaping business operations, industry collaboration, and value generation systems. The market for automotive software updates can generate EUR 20 billion in potential revenue ~\cite{Bazzi2024NovelSystems} while the SDV ecosystem is estimated to achieve a USD 110 billion value by 2030~\cite{BupalamPrasanna2023SystemVehicles}. This market presents multiple opportunities that require collaboration across sectors. One promising approach for developing SDVs is for OEMs to adopt open standards and leverage open-source software (OSS), allowing them to focus their in-house resources on innovation and features that differentiate their brands \cite{khamis2025rethinking}. Prior research on OSS ecosystems shows that competing firms can successfully contribute to the same open-source projects alongside their rivals, highlighting the inherently collaborative and reciprocal nature of OSS development \cite{SCHREIBER2026112765}. An example of such a new initiative is Eclipse Safe Open Vehicle Core (S-CORE), an open-source project within the Eclipse SDV Working Group, aiming to build non-differentiating foundations for various software modules\footnote{https://eclipse.dev/score/}. 

OSS, combined with standardisation functions, can support SDV collaboration by reducing redundant development activities and accelerating innovation through shared platforms. The automotive industry has already adopted open-source solutions, such as Linux and the Robot Operating System (ROS) ~\cite{merlitz2025open}. The industry also benefits from standardised interoperability through the work of consortia, including Eclipse SDV, Automotive Open System Architecture (AUTOSAR), including AUTOSAR Adaptive, COVESA (Connected Vehicle Systems Alliance), and SOAFEE (Scalable Open Architecture for Embedded Edge)~\cite{BupalamPrasanna2023SystemVehicles, Rao2023AcceleratingIndustry, khamis2025rethinking, merlitz2025open}. The open-source and standardisation efforts create network effects, enabling OEMs to create differentiation through software applications rather than building new base architectures, following the Red Hat business model of free open-source distribution with paid support, services, and certification~\cite{merlitz2025open}.

The automotive industry has formed strategic partnerships that combine vehicle expertise with technology from related sectors, transforming supplier relationships. Examples include Ford's collaboration with Google Cloud, Volkswagen's partnership with Microsoft Azure, Mercedes-Benz's integration of NVIDIA's high-performance computing platform, BMW's partnership with Qualcomm for semiconductor solutions, General Motors' joint venture with Cruise for autonomous driving, and Toyota's alliance with Apex AI for software development ~\cite{Rao2023AcceleratingIndustry}. The shift from standard transactional dealings to unified hardware-software solution development becomes evident through Stellantis' direct OEM partnership with Infineon Technologies and Mercedes-Benz's collaboration with NVIDIA~\cite{liebetrau2022architecture}. In turn, system integrators operate as impartial coordinators who use their multi-OEM expertise to create cooperative frameworks for building shared platforms, enabling competitive market diversity~\cite{BupalamPrasanna2023SystemVehicles}. 

However, transformation toward SDVs involves more than forming bilateral partnerships or adopting technological advancements. It requires coordinated adaptation across firms, collaboration, and the broader ecosystem \cite{Paulweber2025Multi-PartnerEcosystem, Peltonen2024Software-DefinedReport}. Historical examples from the mobile industry, such as Nokia, BlackBerry, and Motorola, show how companies can pursue technological advancements and partnerships, and yet fail if they misinterpret ecosystem dynamics, underestimate shifts in control, or neglect the development of critical organisational capabilities \cite{hari2014impact}. These firms were slow to recognise that value creation and innovation increasingly relied on platform-based coordination and collective capabilities instead of just firm-level optimisation. Similar dynamics are now emerging in the automotive industry with the shift to SDVs, as traditional supply-chain logics are challenged. Comparable trends can also be seen in other sectors like digital twin ecosystems, which have evolved from value-chain-orientation towards broader ecosystem collaboration \cite{tripathi2024stakeholders, xu2020digital}. 

Unlike traditional supply chains, which operate through linear interactions and fixed supplier networks, SDVs depend on open, multi-directional collaboration that transcends organisational and domain boundaries \cite{liebetrau2022architecture, khamis2025rethinking}. 
This openness enables the formation of new actors and partnerships around services and products, thereby enhancing adaptability and collective innovation \cite{axelsson2014characteristics}. Thus, ecosystem collaboration in the automotive domain must extend beyond individual partnerships or technological integration, involving multiple interlinked levels from technological and organisational to industry and societal levels. Interactions within and across these levels, such as aligning technical standards, can complicate the design and deployment of SDV applications like autonomous vehicles (AVs) \cite{nickerson2022automated, lusch2014service}. Moreover, shifting power dynamics redefine firms’ roles, creating risks for  dependent actors  while empowering others to influence collaboration \cite{VALENCA2018478}. Understanding these dynamics is crucial, as collective stakeholder efforts shape the balance between innovation and efficiency essential for long-term ecosystem success \cite{Hietala2025GoverningEcosystems}. 

To ensure long-term viability in a multilevel ecosystem collaboration, organisations must align their internal goals and expertise, secure managerial commitment, and leverage external incentives and opportunities to contribute to a shared vision, but achieving this alignment can be challenging \cite{Hietala2025GoverningEcosystems}. As the automotive industry continues to evolve, interdependent challenges, political ambitions, and the increasing demand for continuous updates and mobility services will shape its trajectory \cite{Peltonen2024Software-DefinedReport, EuropeanCommission2025CommunicationSector}. This study aims to clarify the transition from supply chain approaches to ecosystem collaboration, addressing gaps and themes in existing research. It will explore the dynamics of collaboration among stakeholders in these ecosystems.

\section{Methodology}

\begin{figure*}
    \centering
    \includegraphics[width=1\linewidth]{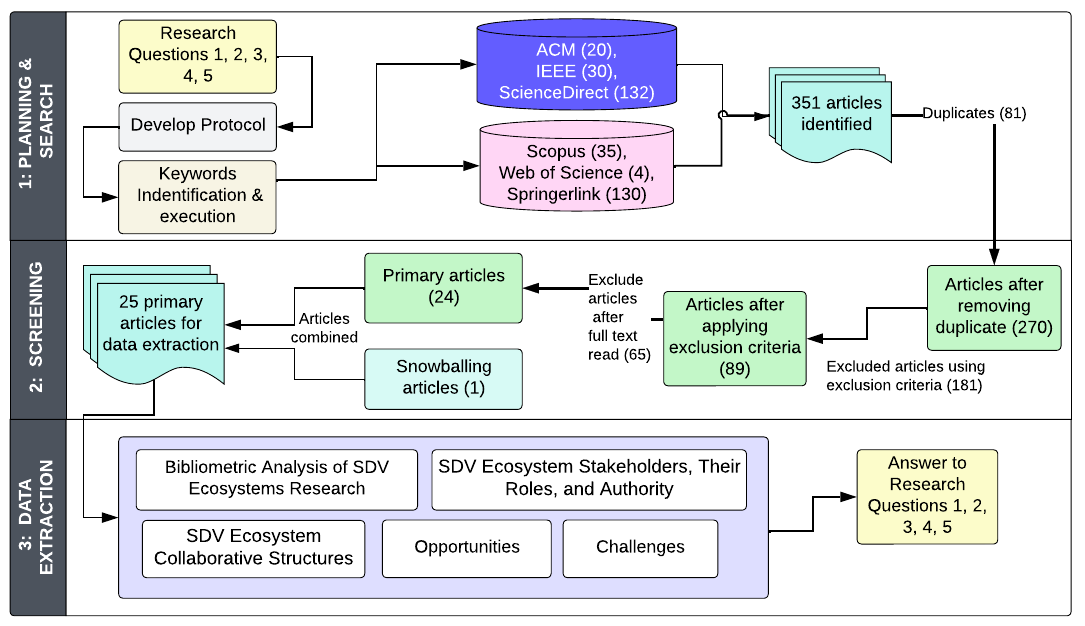}
    \caption{Research Process and SLR Protocol}
    \label{fig:SDVprocess}
\end{figure*}

This study adopts the guidelines by  \cite{Kitchenham2004ProceduresReviews} for conducting SLRs in software engineering. Accordingly, the review process was structured into three main phases: \textit{(i)} planning and search, \textit{(ii)} screening, and \textit{(iii)} data extraction and synthesis. The SLR tool \href{https://parsif.al/}{Parsifal} was utilised to support protocol development, screening, and data management. The overall research process, including search, screening, and analysis steps, is illustrated in Figure~\ref{fig:SDVprocess}.

\subsection{Planning}
\paragraph{Research Objectives and Protocol}

The primary objective of this review is to consolidate current knowledge on \textit{Software-Defined Vehicle (SDV) ecosystems}. More specifically, the review examines how collaborative structures are organised, identifies key stakeholders and their roles, relationships, and decision-making authority, and consolidates the challenges and opportunities shaping SDV-driven industry transformation. The objectives we formulated are based on the research questions summarised in Table~\ref{RQ}.

\paragraph{Protocol Overview}

To address these objectives, a systematic review protocol was developed and iteratively refined by four authors across two phases. The protocol guides the structured identification and analysis of SDV ecosystem dynamics, with a focus on stakeholders, their roles, interactions and interdependencies, and ecosystem-level dynamics. Guided by the Population, Intervention, Comparison, Outcome, and Context (PICOC) framework, the review focuses on populations such as networks and business collaborations, with interventions including ecosystems and industry networks. Our search strategy employs a combination of keywords and synonyms, utilising Boolean operators (AND, OR), to ensure comprehensive coverage. Multiple digital libraries, including ACM, IEEE, ScienceDirect, SpringerLink, Scopus, and Web of Science, are used as primary sources. The selection process applies defined inclusion and exclusion criteria. 

\subsection{Conducting}

\paragraph{Search String and Search}

Search terms were derived from prior SDV reviews \cite{Otto2025TowardsAvenues} and refined through two phases of pilot searches involving two authors. The key terms (Table~\ref{keywords}) reflected both technical and organisational aspects of SDVs, addressing the RQs. As shown in Figure~\ref{fig:SDVprocess}, the initial search identified 351 records across six digital databases (ACM: 20, IEEE: 30, ScienceDirect: 132, Scopus: 35, Web of Science: 4, SpringerLink: 130).

\begin{table}[]
\caption{Search string}
\fontsize{9}{10}\selectfont
\label{keywords}
\begin{tabular}{|p{2cm}|p{6cm}|}

\hline
\textbf{Type}  & \textbf{Search string}                                      \\ \hline

Population (Core SDV terms) & "software vehicle" OR "software car" OR "software defined vehicle" OR "software defined vehiculum" OR "software defined car"  \\  \hline 

Intervention 1 (Ecosystem and network terms) &  “ecosystem”, “industry network”, “innovation network”, “supply network”, “value network”, “collaboration network” \\ \hline 

Intervention 2 (Business and governance terms) & “business model”, “open-source software”, “standardisation”, “cross-industry collaboration”, “value chain”, “supply chain” \\ \hline

\end{tabular}
\end{table}

After removing 81 duplicates, 270 articles remained. Titles, abstracts, and keywords were screened based on predefined exclusion criteria (Table~\ref{criteria}). A total of 89 articles progressed to full-text assessment, after which 65 were excluded due to a lack of alignment with the research objectives. Ultimately, 24 primary studies and one additional paper identified via backward snowballing were included (25 total). These articles are listed in Table \ref{biblio}.

\begin{table}[t] 
\caption{Exclusion criterias, papers excludes and its examples}\label{criteria}
\centering
\begin{tabular}{|m{4.5cm} | m{1.5cm}|  m{1.5cm}|}%
\hline
  \textbf{Exclusion criteria} & \textbf{Frequency} & \textbf{Examples} \\ 
\hline
The paper is not in English. & 11 &  \cite{bhange2023entwicklungsmethoden}, \cite{plagge2023paradigmenwechsel} \\
  \hline
  Paper does not contain keywords or address the main objectives of the study. & 30 & \cite{iida2016ple}, \cite{ghimire2024policy}\\
  \hline
  Retracted article & 1 & \cite{kachhoria2023retracted} \\
  \hline
  Related literature review  & 1 & \cite{khamis2025rethinking} \\
  \hline
  Full text not available/accessible &  17 & \cite{jiang2024vehicle}, \cite{becker2022if} \\
  \hline
  After full text: the paper does not contain keywords or address the main objective. & 65 & \cite{11014909}, \cite{dettinger2024future} \\  
 \hline
  
\end{tabular}


\end{table}

\paragraph{Quality Assessment}

It was conducted using a three-level evaluation (high, medium, low) by three authors in two phases, focusing on methodological rigour, relevance to SDV ecosystems, and contribution to the research questions. The assessment of each article is shown in Table~\ref{biblio}.

\subsection{Analysis and Reporting}

Data were extracted in three phases by three authors using a structured form that captured bibliometric data, collaborative structures, stakeholders, their roles, and authority, as well as opportunities and challenges. Each primary study was coded using the Excel sheet to ensure consistent analysis. A deductive thematic analysis approach was applied, complemented by an inductive analysis \cite{cruzes2011recommended}. The deductive approach categorised data within the predefined themes, while the inductive approach aimed to identify novel insights within those predefined categories. This approach allowed us to synthesise findings that are organised into five analytical dimensions: (1) Bibliometric analysis of SDV ecosystem research, (2) SDV ecosystem collaborative structures, (3) stakeholders and their roles and authority, (4) challenges, and (5) opportunities. An Excel file summarising our data extraction is available in the Appendix supplementary materials. In the next chapter, we will outline the findings of each of these dimensions. 

\subsection{Study Validity and Limitations}

This section addresses potential threats to the validity of our study, categorised into construct, internal, external, and conclusion validity, as noted by Zhou et al. \cite{zhou2016map}. For each category, we outline the associated risks and mitigation strategies employed during the review process.

\paragraph{Construct validity} This concerns whether the SLR procedures accurately capture the concepts addressed in the research questions. Threats to construct validity include insufficient specification, the use of inappropriate or incomplete search terms, the lack of standard terminology, and poorly formulated research questions \cite{zhou2016map}.

To mitigate these risks, we developed the SLR protocol through multiple iterations involving four authors across two phases. The protocol clearly specifies the research objectives, PICOC criteria, search strings, databases, and inclusion/exclusion criteria. We refined the search terms through two rounds of pilot searches, drawing on existing literature on SDVs as well as domain-specific terminology. The PICOC framework also guided the development of clear and focused research questions, minimising the risk of conceptual misalignment. However, despite these efforts, our study has limitations as the terminology surrounding SDVs is rapidly evolving. Consequently, the risk of missing studies due to terminology variations across disciplines or the introduction of new terms cannot be fully eliminated.

\paragraph{Internal validity}This concerns whether the procedures of an SLR are conducted in a way that minimises bias and ensures that the findings are not distorted by the review process. These threats can be, e.g., errors in identifying relevant primary studies, biases in study selection and data extraction, and subjective interpretations \cite{zhou2016map}.

To mitigate these risks, the SLR protocol was developed iteratively, and search terms were refined through two pilot rounds. Quality assessment and data extraction from the primary studies were conducted independently by three authors. Discrepancies were discussed and synthesised collectively to ensure consistent interpretation and minimise individual bias.
Despite these measures, some limitations persist as interpretive judgments cannot be fully avoided when qualitatively synthesising heterogeneous literature.

\paragraph{External validity} This concerns the generalisability of findings within the broader landscape of SDV research. Potential threats to generalisability include a limited time span, restricted venues or databases, and issues related to the generalisability of primary studies \cite{zhou2016map}.

To address these threats, we included six major digital libraries: ACM, IEEE, ScienceDirect, SpringerLink, Scopus, and Web of Science and used backwards snowballing to enhance coverage. Our search strategy and inclusion criteria were specifically designed to capture relevant research across technical, organisational, and ecosystem-oriented domains of SDV. However, one limitation on external validity remains: our review excludes grey literature, such as technical standards, industry white papers, and reports, which are significant in the SDV field. While this exclusion was necessary to maintain methodological rigour, it does limit the generalisability of our findings.

\paragraph{Conclusion validity} It addresses whether analytical procedures yield reliable and replicable conclusions. Threats include, e.g., subjective interpretations of extracted data, inadequate data synthesis, and the impact of issues in primary studies on the accuracy of conclusions \cite{zhou2016map}.

To address these concerns, all extracted data were cross-checked by three authors. Any disagreements were resolved through discussions until a consensus was achieved. We employed a robust deductive and inductive thematic analysis in our data analysis \cite{cruzes2011recommended}. Deductive coding followed predefined categories (such as ecosystem structures and stakeholder roles), while inductive coding sought to identify new themes. Despite these efforts, the conclusions remain dependent on the quality, depth, and reporting practices of the primary studies. Given the novelty of SDV research, the literature is fragmented and empirically varied, which may limit the robustness of the conclusions.

\section{Results}

\subsection{Bibliometric Analysis of SDV Ecosystems Research (RQ1)}

\begin{figure}[h!]
    \centering
    \includegraphics[width=1\linewidth]{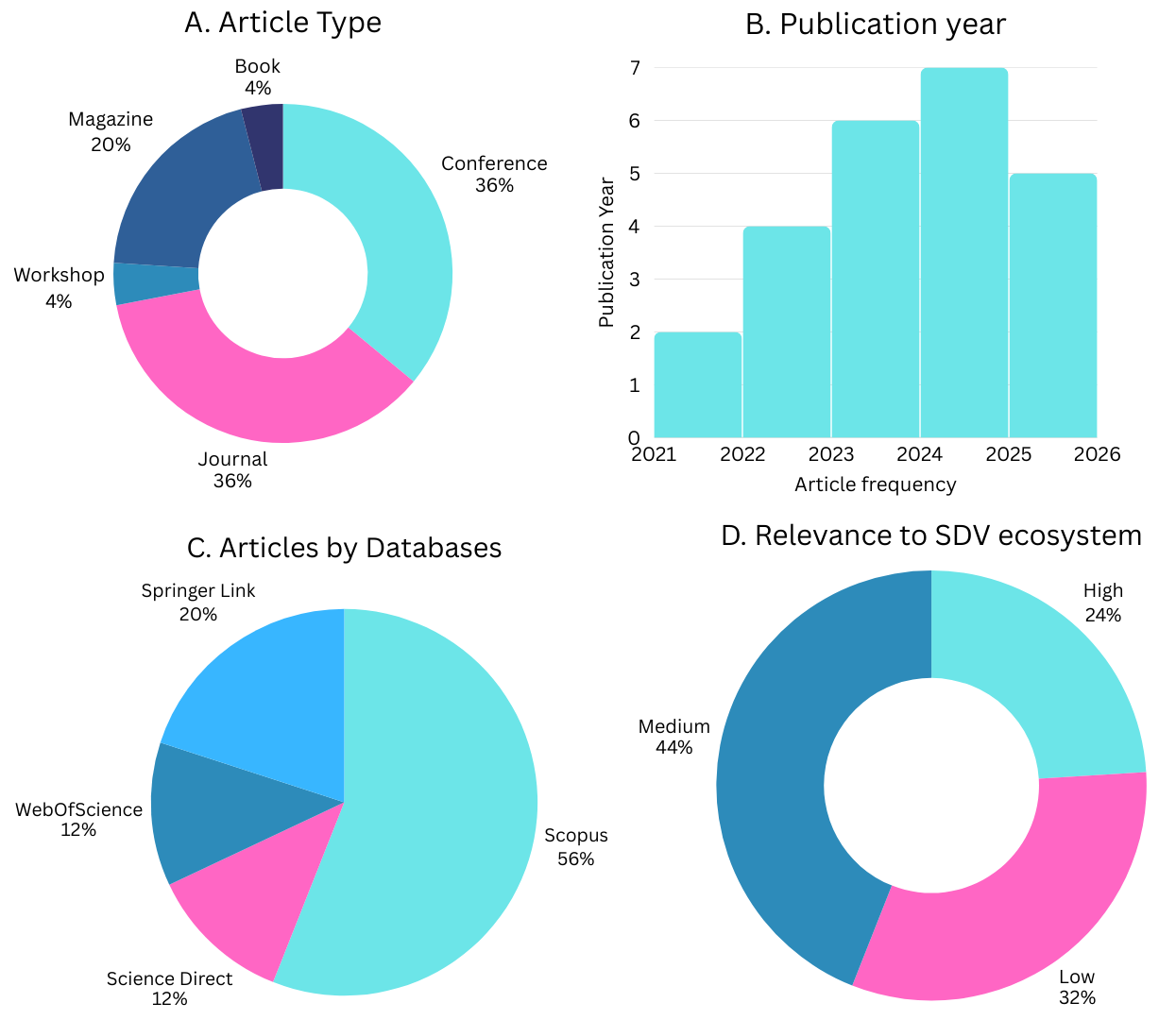}
    \caption{Bibliometric analysis overview}
    \label{Bibliometric}
\end{figure}

This SLR consolidates research from 25 primary studies published between 2021-2025 (Table \ref{biblio}). As can be seen in Figure \ref{Bibliometric}, the studies included one book and the rest were published in a workshop (1), conferences (9), magazines (5), and journals (9). Publication activity has increased steadily since 2021, reflecting the growing interest in industrial and academic collaboration around SDVs and their ecosystem implications. Although the studies cover a relatively short time frame, together they offer an overview of current trends and advancements in SDV ecosystems.

The primary studies were mostly published in automotive and general engineering venues, with limited representation from software engineering, information systems and management journals. As a result, the research remains largely technical, focusing on the standardisation of the SDV architecture, interoperability, and cybersecurity, while leaving ecosystem governance, particularly in terms of collaboration around software, business models, and regional variation, to receive less attention. European perspectives, including those from the German automotive industry, emerge (5) alongside growing contributions from Asian contexts, specifically China, Japan, and Taiwan (2). In contrast, research focusing on the U.S. is limited, despite U.S. OEMs being frequently referenced in primary articles. Overall, the literature draws on global industry examples while discussing SDV- and ecosystem-related collaborations at a general conceptual level.

While current research provides a solid starting point for understanding software and organisational aspects of SDV ecosystems, empirical studies on governance structures, business models, value distribution, and long-term sustainability remain scarce.

\begin{table*}[h!]
\caption{Primary Articles}
\fontsize{9}{10}\selectfont
\label{biblio}
\begin{tabular}{|p{0.5cm}|p{1.2cm}|p{1,4cm}|p{0.7cm}| p{1.4cm}| p{4,9cm}| p{2cm}| p{2.7cm}|}

\hline
\textbf{ID} & \textbf{Primary Article} & \textbf{Relevance} & \textbf{Year} & \textbf{Article Type} & \textbf{Publication Forum} & \textbf{Database} & \textbf{Geographical Focus}  \\ \hline

P1	& \cite{Narla2025FrameworkShells}    & Low   &  2025	    &  	Conference &    International Conference on Industrial Cyber-Physical Systems     & IEEE & No specific focus	\\  \hline 
P2	& \cite{Bazzi2024NovelSystems}       & Medium  &  2024       &  	Journal &   IEEE Access   & IEEE, Scopus & No specific focus (global examples) \\  \hline 
P3	& \cite{prasanna2023system}          & High  &  2023	    &  	Conference  &   IEE International Transportation Electrification Conference    & Scopus &	No specific focus\\  \hline 
P4	& \cite{Rao2023AcceleratingIndustry} & High  &  2023	    &  	Conference  &	IEEE International Transportation Electrification Conference     & Scopus & No specific focus (global examples) \\  \hline 
P5	& \cite{robinson2024ai}              & Low   &  2024	    &  	Conference  &	International Conference on Electrical and Computer Engineering Researches      & Scopus & No specific focus (global examples)  \\  \hline 
P6	& \cite{Parkhurst2025AutomatedDevelopment}  & Low  & 2025	&  Journal     &  Transportation Research Procedia      &	Science Direct & UK \\  \hline 
P7	& \cite{Panchal2023DesignOpportunities}  & High    &  2023	&  Journal      &  Journal of Computing and Information Science in Engineering     & Scopus & No specific focus	\\  \hline 
P8	& \cite{wu2021display}               & Medium    & 	2021	&  	Workshop    &  International Workshop on Active-Matrix Flatpanel Displays and Devices    & IEEE	& Asia (China, Taiwan, Korea, and Japan) \\  \hline 
P9	& \cite{liebetrau2022architecture}   & Medium    &  2022	&  	Conference  &   Automotive meets Electronics   & IEEE, Scopus 	& No specific focus\\  \hline 
P10	& \cite{Liu2022ImpactVehicles}       & High     &  	2022	&  	Journal     &	Automotive Innovation     & Scopus, Web of Science & No specific focus \\  \hline 
P11	& \cite{rehberg2024industrial}       & Low      &  	2024	&  	Journal     &	Journal of Engineering and Technology Management  & Science Direct  & Europe (Germany) \\  \hline 
P12	& \cite{szalavetz2024inhouse}       & High      &  	2024 	&  	Journal     &	International Journal of Automotive Technology and Management    & Scopus & No specific focus (global examples)  \\  \hline 
P13	& \cite{laaschmaximize}             & Low       &  2023	    &  	Magazine    &   ATZelectronics worldwide       & Springer Link & Europe	\\  \hline 
P14	& \cite{Paulweber2025Multi-PartnerEcosystem}  & High  &	2025 &  Conference  &   Design, Automation \& Test in Europe Conference  &	IEEE, Scpus, Web of Science & Europe \\  \hline 
P15	& \cite{merlitz2025open}            & Medium    &	2025	&  	Magazine      &	 ATZelectronics worldwide    & Snowballed &  No specific focus\\  \hline 
P16	& \cite{becker2022operating}        & Medium    &  	2022	&  	Magazine      &  ATZelectronics worldwide   & Springer Link &	No specific focus\\  \hline 
P17	& \cite{spencer2023real}            & Low       &  	2023	&  	Conference   &	Cyber-Physical Systems and Internet of Things Week      & Scopus & No specific focus \\  \hline 
P18	& \cite{tang2025research}           & Medium    &  	2025	&  	Journal      &  Systems   & Scopus, Web of Science  & China	\\  \hline 
P19	& \cite{unseldshaping}              & Medium    &  	2021	&  	Magazine     & ATZelectronics worldwide   & Springer Link &	No specific focus\\  \hline 
P20	& \cite{unseld2025significant}      & Low       &  	2025	&  	Magazine     &  ATZelectronics worldwide	  & Springer Link &  Europe (Germany)   \\  \hline 
P21	& \cite{bhandari2024software}       & Medium    &  	2024	&  	Conference   &  International Conference on ICT for Sustainable Development  & Scopus	& No specific focus\\  \hline 
P22	& \cite{aust2022vehicle}            & Low       &  	2022	&  	Conference   &  International Symposium on Wireless Personal Multimedia Communications    & IEEE, Scopus &   No specific focus	\\  \hline 
P23	& \cite{lu2024vehicle}              & Medium    &  	2024	&  	Book         &	Publisher: Springer Cham   & Scopus & No specific focus (global examples) \\  \hline 
P24	& \cite{lu2023vehicle}              & Medium    &  	2023	&  	Journal      & Journal of Information and Intelligence   & Science Direct &  No specific focus (global examples) \\  \hline 
P25	& \cite{wu2024vpi}                  & Medium    &  	2024	&  	Journal      &	Journal of Computer Science and Technology   & Springer Link & No specific focus (global examples) \\  \hline 

\end{tabular}
\end{table*}

\subsection{SDV Ecosystem Collaborative Structures (RQ2)}

\begin{table*}[h!]
\caption{Multiple Levels of Collaborative Structures in SDV Ecosystems}
\fontsize{9}{10}\selectfont
\label{structures}
\begin{tabular}{|p{2.5cm}|p{4.3cm}|p{7cm}| p{2.7cm}|}

\hline
\textbf{Level of Collaboration} & \textbf{Stakeholders Involved}   & \textbf{Rationale / Explanation} & \textbf{Source} \\ \hline

Intra-organisational collaboration	&	OEMs, in-house software teams, internal software subsidiaries	&	OEMs establish internal software units to manage SDV architectures and integration, reducing reliance on external suppliers. &	P2, P4, P5, P7, P9, P10, P12, P13, P15, P16, P21 \\  \hline 
Bilateral/strategic partnerships	&	OEMs and technology firms (e.g., Ford–Google, Toyota–AWS, Volkswagen–Microsoft, Baidu–Geely)	&	Partnerships combine automotive and digital expertise to develop AI, cloud-based, and data-driven mobility platforms.	& P4, P5, P7, P8, P9, P12, P13, P18, P19, P20, P21, P23, P24, P25 \\  \hline 
Supplier and value-chain collaboration	&	OEMs, Tier-1–3 suppliers, semiconductor firms, software firms	&	E/E transformation creates inter-tier dependencies, requiring collaboration on hardware–software integration and safety compliance.	& P2, P3, P4, P7, P8, P9, P10, P11, P12, P13, P14, P19, P21, P24, P25 \\  \hline 
Open-source collaboration	&	Open-source communities, OEMs, Tier suppliers, software firms	&	Initiatives like Eclipse OSS SDV, AGL, OpenADX, ROS, and IVY enable shared code development, modularity, and cost-efficient system integration.	& P2, P3, P4, P13, P14, P15, P16, P17, P19, P20, P22, P24, P25 \\  \hline 
Industry consortia and standardisation networks	&	OEMs, suppliers, software developers, infrastructure providers, regulators	&	Initiatives such as AUTOSAR, COVESA, and SOAFEE  coordinate standardisation for interoperability, cloud-based computing, and safety (ISO 26262).	& P1, P2, P3, P4, P7, P14, P15, P16, P17, P18, P19, P20, P21, P22, P25 \\  \hline 
Policy-, research-driven, and cross-sectoral collaboration	&	Governments, OEMs, suppliers, research institutions, EU initiatives (e.g., FEDERATE, SDV Sherpa Group, Autotech.agil)	&	Public–private programmes coordinate non-differentiative SDV components, shared architectures, and regional alignment on SDV innovation.	& P3, P4, P5 P6, P7, P10, P11, P14, P18, P20, P21, P23, P24, P25  \\  \hline

\end{tabular}
\end{table*}

The transition toward SDVs is reshaping the automotive industry, which now faces rising costs, rapid technological change, intensifying global competition, and growing consumer expectations [P14, P15, P21]. Technological development toward SDVs both drives and demands ecosystem collaboration, as challenges related to interoperability, data management, and system integration necessitate cooperation among multiple software vendors and partners [P1–P3, P7]. This transformation affects the entire value chain, pushing suppliers to rethink their business models and move away from linear chains toward interdependent networks and new revenue logics such as optimisation and servitisation [P3, P9]. Yet, differing strategic choices among OEMs and suppliers, whether to outsource, collaborate, or rely on in-house R\&D, often create conflicting goals and power tensions within the ecosystems [P10]. 

While the automotive industry has traditionally operated through hierarchical structures with long lead times [P8], our findings show that SDV ecosystems are instead organised around multiple layers of collaboration, where established and emerging actors jointly address the technical and organisational demands of software-centric mobility.  Table \ref{structures} summarises six levels of collaborative structures identified in the primary studies. 

OEMs are increasingly internalising software expertise [P12, P16], while strategic partnerships with major technology firms, such as Ford–Google, Toyota–AWS, and Baidu–Geely, bridge automotive engineering with cloud, AI, and data capabilities [P4, P23, P24]. Collaboration along the supply chain now spans from Tier 1 to Tier 3 suppliers and semiconductor firms to address hardware–software integration and safety compliance [P21]. Industry consortia and standardisation networks such as AUTOSAR, COVESA, and SOAFEE promote interoperability and standardised architectures [P2, P7]. Open-source initiatives, such as Eclipse OSS SDV, AGL, OpenADX, ROS, and BlackBerry IVY, enable modular and cost-efficient system integration [P14, P15, P24], while policy- and research-driven programs, including FEDERATE and the SDV Sherpa Group, align public and private efforts toward non-differentiating components and sustainable SDV innovation [P14]. Overall, hierarchical supply chains are giving way to networked collaborations that connect OEMs, suppliers, software developers, infrastructure providers, researchers and regulators [P9, P10, P23]. 

\subsection{Stakeholders, Their Roles, and Authority (RQ3)}

\subsubsection{Stakeholders}

The primary articles have examined various stakeholders relevant to SDV ecosystems. These stakeholders are categorised and summarised in Table \ref{stakeholders}, which organises them into 12 distinct stakeholder groups. Key stakeholders include traditional automotive OEMs such as Mercedes-Benz, Toyota, and General Motors (GM), alongside newer players like Tesla and BYD. The studies also feature a wide range of tier companies, software developers, and those involved in software platforms and tools. Stakeholders also include semiconductor manufacturers, big tech firms, various open-source initiatives, industry consortia, government entities, regulatory bodies, and research institutions.

\begin{table*}[h!]
\caption{Stakeholders and Stakeholder Groups}
\fontsize{9}{10}\selectfont
\label{stakeholders}
\begin{tabular}{|p{2cm}|p{5.7cm}|p{5.6cm}| p{3.2cm}|}

\hline
\textbf{Stakeholder Group} & \textbf{Description} & \textbf{Examples from the primary articles} & \textbf{Source} \\ \hline

Traditional Automotive OEMs	&	Established automotive OEMs transitioning from hardware-centric to software-driven architectures; they face challenges integrating SDV solutions into legacy systems.	&	Volkswagen, Toyota, GM, Ford Motor Company, Mercedes-Benz, BMW, Stellantis, Geely Auto, Volvo Cars, SAIC Motor, Honda	& P1, P2, P3, P4, P5, P7, P9, P10, P11, P12, P14, P15, P16, P17, P18, P19, P20, P21, P22, P23, P24, P25 \\  \hline 
New / EV-oriented Automotive OEMs	&	Recently established automotive OEMs, often founded around electric and software-based mobility concepts; they adopt centralised architectures and agile development models.	&	Tesla, NIO, BYD	& P4, P5, P8, P13, P23, P24 \\  \hline 
Tier 1–3 Suppliers and Component Providers	&	Supply core vehicle components (e.g., sensors, batteries, displays) and embedded systems; key partners for OEMs in hardware and integration layers.	&	Bosch, Aptiv, STMicroelectronics, TSMC, AUO, Innolux, LG Display, LG Electronics, Infineon	& P4, P8, P9, P10, P13, P14, P15, P19, P21, P23, P24  \\  \hline 
Software and Platform Developers	&	Develop automotive operating systems, middleware, AI platforms, and domain-specific software that enable SDV operations.	&	CARIAD, Etas, Escrypt, Apex.AI, Green Hills Software, Renovo Motors, Carmera, Hexad, Intenta GmbH, AiMotive, Zenseact, Cruise Automation, ALGOLiON, AUTOSAR, BlackBerry	& P2, P4, P9, P12, P13, P14, P15, P16, P19, P21 \\  \hline 
Semiconductor and Hardware Manufacturers	&	Provide microchips, processors, and computing platforms that power SDV functionalities such as AI, connectivity, and autonomy.	&	Ericsson, NVIDIA, STMicroelectronics, TSMC, Qualcomm & P4, P9, P12, P14, P24, P25	\\  \hline 
Technology and Cloud Companies / Big Tech	&	Deliver cloud computing, AI services, and data infrastructure; drive digital transformation and connectivity in vehicles.	&	Amazon Web Services (AWS), Microsoft (Azure), Google (Google Cloud), Alibaba, Baidu, Huawei, Tencent, BlackBerry	& P4, P7, P12, P17, P18 P23, P24, P25  \\  \hline 
Open-Source Initiatives and Industry Consortia	&	Foster collaboration, standardisation, and interoperability across the automotive software ecosystem.	&	COVESA, AUTOSAR, Eclipse Foundation, SOAFEE, Autoware, ROS, Baidu Apollo, BlackBerry IVY	& P7, P9, P10, P14, P15, P16, P17, P22, P25 \\  \hline 
Mobility and Service Providers	&	Deliver connected vehicle–based mobility services and integrate SDVs into broader digital ecosystems.	&	Cavnue, connected vehicle service enterprises, infrastructure developers, smart home integrators	& P12, P14, P23, P24 \\  \hline 
Government and Regulatory Bodies	&	Establish policies, safety standards, and regulatory frameworks for SDVs and autonomous mobility.	&	UNECE, German Motor Transport Authority, national and urban authorities	& P6, P8, P10, P11, P14, P18, P20, P23, P24 \\  \hline 
Research and Academic Institutions	&	Conduct research and innovation on SDV technologies, standards, and ecosystem collaboration.	&	Universities, scientific research institutions, cross-disciplinary R\&D teams	& P14, P18, P22 \\  \hline 
Financial and Insurance Institutions	&	Fund automotive and SDV innovation, manage investment risk, and develop insurance models for data-driven mobility.	&	Financial institutions, insurance companies, investment organizations	& P18, P22, P23, P24  \\  \hline 
Customers and End Users	&	Use SDVs and services, providing data and feedback that influence system design and business models.	&	Vehicle owners, drivers, passengers, general public, pedestrians, cyclists	& P3, P6, P7, P12, P16, P18, P23, P24 \\  \hline

\end{tabular}
\end{table*}

\subsubsection{Roles of the Stakeholders}

The 12 stakeholder groups were further classified as primary, secondary, and tertiary stakeholders based on their roles in the ecosystem (Table \ref{roles}).

\begin{table*}[h!]
\caption{Stakeholder Hierarchy and Overall Roles}
\fontsize{9}{10}\selectfont
\label{roles}
\begin{tabular}{|p{1.2cm}|p{6.4cm}|p{6.2cm}| p{2.8cm}|}

\hline
\textbf{Level} & \textbf{Stakeholder Groups}   & \textbf{Rationale / Explanation} & \textbf{Source} \\ \hline

Primary Stakeholders & 
Traditional Automotive OEMs

New / EV-oriented Automotive OEMs

Tier 1–3 Suppliers and Component Providers

Software \& Platform Developers & These stakeholders are directly responsible for developing, manufacturing, and integrating SDVs. They form the core value-creation network, where SDV architectures, operating systems, and vehicle platforms are built. & P1, P2, P3, P4, P5, P7, P8, P9, P10, P11, P12, P13, P14, P15, P16, P17, P18, P19, P20, P21, P22, P23, P24, P25 \\ \hline

Secondary Stakeholders & Semiconductor \& Hardware Manufacturers

Technology \& Cloud Companies / Big Tech

Open-Source Initiatives \& Industry Consortia

Mobility \& Service Providers
& These stakeholders enable and support SDV development through technology, infrastructure, collaboration frameworks, and platform ecosystems. They expand the digital and operational backbone of the SDV ecosystem. & P4, P7, P9, P10, P12, P14, P15, P16, P17, P18, P22, P23, P24, P25 \\ \hline

Tertiary Stakeholders & Government \& Regulatory Bodies

Research \& Academic Institutions

Financial \& Insurance Institutions

Customers \& End Users
& These stakeholders influence, fund, and govern the ecosystem rather than directly producing SDV technologies. They provide the institutional, financial, and societal context within which SDV ecosystems evolve. & P3, P6, P7, P10, P11, P12, P14, P16, P18, P20, P22, P23, P24 \\ \hline

\end{tabular}
\end{table*}

\paragraph{Primary stakeholder group} Automotive OEMs, both traditional and new, are key orchestrators in SDV ecosystems, responsible for developing and operating vehicle platforms that integrate software, cloud connectivity, and hardware into cohesive architectures [P4, P7, P10]. Many are internalising software development, including system and application software [P9, P12, P16] while collaborating directly with Tiers [P13], system-on-chip (SoC) suppliers and defining hardware specifications for electronic manufacturing service (EMS) providers [P9]. Strategic partnerships with technology firms like NVIDIA, Google Cloud, Microsoft Azure, and AWS enable incorporating of AI, cloud-based services, and digital twin environments that connect internal and external stakeholders [P18, P23, P24]. Traditional OEMs are shifting from mechanical system integrators to software mobility firms, expanding their revenue streams through software-enabled features, updates, and post-sale digital services [P10, P12].

Tier 1–3 suppliers and component providers play an evolving role as OEMs assume more software responsibilities. Tier 1 suppliers focus on hardware standardisation, integration, and modular architectures that support software scalability, while Tier 2 and Tier 3 suppliers, like semiconductor and component manufacturers, work directly with OEMs to deliver high-performance components for increasingly centralised E/E systems [P9, P13, P14, P19, P21]. This collaboration is shifting from traditional linear supply chains to networked and multi-tiered partnerships, focused on transparency and innovation [P13]. Panel and electronics suppliers (e.g., AUO, Innolux, LG Display, LG Electronics) advanced display technologies, while regional suppliers in Asia strengthen the supply base with cost-efficient, high-quality components [P8].

Software and platform developers provide the digital infrastructure that enables SDV functionality. Companies such as CARIAD, Apex.AI, ETAS, Escrypt, AiMotive, and Green Hills Software aid OEMs and Tier 1 suppliers in developing middleware, abstraction layers, and testing frameworks that ensure modularity, performance, and compliance with standards [P4, P12, P13]. They integrate safety-critical systems, ADAS, and infotainment with frameworks from open-source initiatives such as AUTOSAR, COVESA, and SOAFEE [P17, P19, P21]. These developers are increasingly involved in joint ventures and collaborations across industries that combine elements of automotive, AI, and cloud technology [P2, P23, P24]. This positions them as contributors to the ongoing shift towards more cloud-based and adaptable vehicles.

\paragraph{Secondary stakeholder group} Secondary stakeholders provide the enabling technologies, infrastructures, and collaborative frameworks that allow software-defined vehicles to operate as digital, connected systems. Semiconductor and hardware manufacturers such as NVIDIA, STMicroelectronics, and TSMC supply high-performance computing chips, AI accelerators, and SoCs that underpin autonomous driving, ADAS, and centralised E/E architectures of the future [P4, P9]. Technology and cloud companies, including AWS, Microsoft Azure, Google Cloud, and Alibaba Cloud, deliver the cloud-based infrastructure and data platforms that enable real-time analytics, OTA updates, and the operation of digital twin environments [P23, P24, P25]. Their AI-driven and data-centric services support OEMs in accelerating feature delivery and maintaining software-defined platforms across multi-brand ecosystems. Open-source initiatives and industry consortia, such as AUTOSAR, COVESA, SOAFEE, and Eclipse, coordinate collaboration by developing shared standards, APIs, and modular software architectures [P14, P15]. They facilitate open innovation and interoperability across diverse systems, helping to align OEMs, suppliers, and software developers around common frameworks for safety-critical, cloud-native automotive systems[P10, P17]. Finally, mobility and service providers, including infrastructure developers like Cavnue, extend the SDV ecosystem into operational contexts. Their role is to deliver data-driven and networked mobility solutions that integrate vehicles with smart infrastructure, traffic management, and digital service environments [P23, P24]. Together, these secondary stakeholders have a supportive role in transforming the SDV landscape into an interconnected ecosystem.

\paragraph{Tertiary stakeholder group} Tertiary stakeholders shape the development and operational environment of SDV ecosystems. Government and regulatory bodies, including national administrations, urban authorities, and international organisations such as UNECE, establish the framework for SDV development through legislation and compliance frameworks [P6, P11, P20]. Their role is to enforce laws and regulations (e.g., UNECE R155 and R156 for cybersecurity and software update management), set incentives, promote industrial collaboration, and establish common technical standards to ensure safety, interoperability, and accountability across borders [P8, P11, P20]. Research and academic institutions contribute to advancing the technological and methodological foundations of SDVs by studying cutting-edge topics, such as the edge-cloud continuum, AI-driven mobility, and ecosystem governance [P14, P18, P22]. They work in joint projects and cross-disciplinary R\&D collaborations like FEDERATE, and autotech.agil that integrate hardware, software, and data-driven innovation [P14, P20]. Financial and insurance institutions underpin the economic viability of SDV ecosystems by funding technological development, managing investment risk, and designing new insurance models for data-driven, connected mobility [P18, P23, P24]. Ultimately, customers and end-users influence SDV design and market direction through their expectations for integrated, smartphone-like functionalities, seamless digital experiences, and continuous software updates [P12, P16, P18]. Their usage patterns and feedback drive OEMs and service providers to align innovation with user value, safety, and trust, thereby creating an active role for end-users in the SDV ecosystem, rather than casting them as passive recipients of technology [P12, P23, P24].

\subsubsection{Authority - Power, Control and Decision-making}

\begin{table*}[h!]
\caption{Power, Control, and Decision-Making}
\fontsize{9}{10}\selectfont
\label{power}
\begin{tabular}{|p{2.2cm}|p{12.8cm}|p{2.1cm}|}

\hline
\textbf{Theme} & \textbf{Description} & \textbf{Source} \\ \hline

Centralisation of control by OEMs	&	OEMs maintain and even expand control over system structure, data exchange, and software integration by taking responsibility for software, coordinating with suppliers, defining hardware, and acquiring software firms.	& P1, P2, P9, P12, P13, P16 \\  \hline 
Redistribution of power along the value chain	&	The shift from vertical supply chains to networked ecosystems redistributes power among OEMs, suppliers, and software firms, intensifying competition for dominance. OEMs are increasingly collaborating directly with tiered suppliers, while software companies gain leverage through their expertise and critical inputs. & P3, P9, P10, P13, P14, P18, P20, P21, P23, P24	\\  \hline 
Platformisation and the emergence of digital gatekeepers	&	The industry is shifting to a platform-based ecosystem where OEMs and platform firms define vehicle architectures. Big Tech companies like Apple and Google can gain influence by controlling digital interfaces and cloud services, potentially becoming gatekeepers. OEMs strive for autonomy but often integrate these external systems to meet customer demands. & P7, P10, P12, P16, P23	\\  \hline 
Control–openness trade-off	&	OEMs and automotive supply chains face a choice between developing proprietary solutions for autonomy or adopting open standards for better interoperability. In-house development offers control but requires significant resources, while open source boosts collaboration and integration speed. Open-source software is vital for addressing technological and geopolitical challenges by encouraging standardised modular architectures, Software Development Kits (SDKs), and collective innovation. & P3, P12, P14, P15, P16, P17, P19, P20	\\  \hline 
Regulatory, regional, and political dimensions of power	&	Governments and international bodies enforce laws that require companies to comply, influencing control and accountability in the SDV ecosystem. Power dynamics differ regionally: Chinese suppliers leverage state support and domestic demand, while European players collaborate through initiatives like FEDERATE, uniting OEMs and other stakeholders to establish standards and conduct research. National and regional politics impact the pace and direction of SDV adoption. &	P5, P8, P11, P14, P18, P20 \\  \hline

\end{tabular}
\end{table*}

Cross-cutting themes from the primary articles highlight the distribution of power, control, and decision-making in SDV ecosystems and their impact on collaboration, efficiency, and innovation. Table \ref{power} summarises these themes.

OEMs strengthen their authority by defining system architectures, managing data exchange, and internalising software development through acquisitions and dedicated divisions [P1, P2, P12, P13]. While newer SDV-focused firms adapt quickly, traditional OEMs must reconcile legacy processes with agile software practices [P7]. 

The shift from vertical supply chains to networked ecosystems redistributes authority, giving software and technology companies greater influence due to their technical expertise and dominance in other ecosystems [P3, P10, P21]. As the industry evolves into a platform-based ecosystem, OEMs and Big Tech firms are not only competing for control over digital interfaces and cloud infrastructures but are also forming strategic partnerships to gain a competitive power [P12, P16]. Moreover, those in power face a strategic choice between proprietary systems that secure autonomy and open, standardised solutions that foster interoperability and faster innovation [P3, P12, P15, P16]. Meanwhile, governments and international bodies shape compliance and coordination through regulations and incentives, with regional and political factors also influencing these dynamics [P5, P18, P20]. For example, Chinese suppliers benefit from state support and high local demand, whereas European collaboration emphasises shared governance and standardised development, with initiatives like FEDERATE promoting collaboration, joint decision-making and standardisation [P8, P14]. 

Overall, the distribution of power, control and decision-making within SDV ecosystems is shaped by organisational structures, technological capabilities, and regional contexts, resulting in complex dynamics that affect collaboration, efficiency, and innovation.

\subsection{Challenges (RQ4)}

The primary articles address a variety of challenges that can be grouped into four themes, including (1) software, (2) organisational, (3) industry and market, and (4) regulatory, legal, and ethical challenges.

\textbf{Software challenges} (Table \ref{software challenges}) in SDV ecosystems involve managing the increasing complexity, interoperability, and security demands of SDVs. The integration of multiple software modules, ECUs, and third-party systems remains difficult due to inconsistent architectures, fragmented standards, and vendor-specific interfaces [P2, P11, P23]. Limited standardisation, such as underdeveloped middleware layers, hampers cross-platform compatibility and slows down the adoption of open-source solutions that could otherwise reduce costs and enhance scalability [P10, P15]. Cybersecurity poses a persistent threat as vehicles expose multiple attack surfaces through various interfaces, cloud connections, and OTA updates, requiring multilayer protection, encryption, and continuous monitoring [P4]. The decoupling of software from hardware further complicates development, demanding standardised communication protocols and lifecycle synchronisation between physical components and digital services [P7]. Managing vast data flows, real-time performance, and safety-critical software integration amplifies these difficulties, particularly in autonomous systems, which involve hundreds of interdependent processes [P24]. Compounding these technical issues, rigid architectures, high maintenance costs, and shortages of skilled developers limit the industry’s ability to implement modular, updatable, and secure software frameworks at scale [P3, P13].

\begin{table*}[h]
\caption{Software development Challenges}
\fontsize{9}{10}\selectfont
\label{software challenges}
\begin{tabular}{|p{3cm}| p{12.5cm}|p{1.5cm}|}

\hline
\textbf{Category} & \textbf{Description} & \textbf{Source} \\ \hline

Integration and interoperability	&	Integrating multiple software modules, ECUs, and third-party systems remains difficult due to fragmented architectures, incompatible standards, and underdeveloped middleware layers. Ensuring seamless data flow, real-time communication, and interoperability across domains is essential yet unresolved.	&	P1, P2, P11, P14, P16, P21	\\  \hline 
Standardisation and open-source limitations	&	Lack of unified standards, reliance on proprietary systems, and limited adoption of open-source solutions hinder scalability and collaboration. Limited availability of open vehicular platforms and immature SOA restrict flexibility and cross-industry development.	&	P1, P10, P15, P23	\\  \hline 
Cybersecurity and data protection	&	Expanding connectivity exposes multiple attack surfaces, from OBD and wireless interfaces to cloud systems. Maintaining confidentiality, integrity, and availability requires multilayered security, encryption, and secure OTA updates.	&	P2, P4, P21	\\  \hline 
Software–hardware decoupling and compatibility	&	Aligning hardware and software lifecycles is challenging, as decoupled architectures demand standardised interfaces and hardware designed for continuous software-driven upgrades.	&	P7, P22	\\  \hline 
Complexity and scalability management	&	Rapidly increasing software size and real-time data demands require modular, scalable architectures that safely integrate safety- and non-safety-critical components while balancing cost and performance.	&	P5, P16, P19, P24, P25	\\  \hline 
Maintenance and lifecycle limitations	&	Custom hardware, and rigid OEM architectures increase ownership costs, limit serviceability, and reduce accessibility for third-party integration and long-term digital transformation.	&	P3, P13, P23	\\  \hline 

\end{tabular}
\end{table*}

\textbf{Organisational challenges} (Table \ref{org challenges}) in SDV ecosystems stem from the industry’s structural, cultural, and capability transitions, as it shifts from mechanical manufacturing to software-driven innovation. Traditional, hierarchical structures and control-oriented engineering cultures struggle to adapt to agile, iterative development models and cross-functional collaboration [P3, P7]. Talent shortages and limited cross-disciplinary expertise hinder integration across software, hardware, and cloud domains, slowing development and innovation [P10]. Coordinating complex workflows across diverse suppliers, internal teams, and technology domains exposes inefficiencies, especially in aligning software and hardware lifecycles [P7]. Firms also face strategic uncertainty in balancing internalisation, outsourcing, and open-source participation. Building in-house capabilities demands high investment and expertise, while adopting open or collaborative approaches requires structural and cultural changes [P12, P15, P16]. Uneven organisational capacities exacerbate these issues—new entrants, like BYD or Tesla, advance more rapidly with integrated software cultures, whereas incumbents face costly transformations, legacy constraints, and uneven progress toward software-defined operations [P5, P9].

\begin{table*}[h]
\caption{Organisational Challenges}
\fontsize{9}{10}\selectfont
\label{org challenges}
\begin{tabular}{|p{3cm}| p{12.5cm}|p{1.5cm}|}

\hline
\textbf{Category} & \textbf{Description} & \textbf{Source} \\ \hline
Cultural and structural transformation	&	Traditional automotive OEMs must shift from mechanical and control-oriented engineering toward agile, software-driven practices. Balancing iterative software cultures with safety-focused rigour exposes cultural clashes and legacy constraints.	&	P3, P7, P10, P11, P13, P15	\\  \hline 
Talent and capability gaps	&	There is a shortage of cross-disciplinary talent skilled in both hardware and software. Traditional OEMs struggle to attract and retain skilled developers, particularly in Europe, with high labour costs, and to build efficient, software-focused teams.	&	P4, P9, P10, P11, P13 	\\  \hline 
Integration across disciplines and functions &  Aligning hardware, software, and AI development requires multidisciplinary coordination, new design tools, and integrated workflows across cloud, network, and middleware layers. Legacy structures impede this co-evolution.	&	P2, P7, P24	\\  \hline 
Internalisation versus outsourcing and open-sourcing	&	OEMs face high risks and inefficiencies when internalising software capabilities without adequate expertise, while outsourcing and open-sourcing can limit autonomy. Strategic balance between proprietary development and ecosystem collaboration remains a challenge.	&	P10, P12, P15, P16	\\  \hline 
Collaboration and coordination barriers	&	Complex value chain transformation demands sustained coordination among OEMs, suppliers, and partners. Limited expertise, resource constraints, and inconsistent cooperation hinder co-creation and innovation.	&	P1, P18, P21, P23  	\\  \hline 
Organisational capacity and uneven progress	&	Readiness levels vary significantly: companies like Tesla have advanced AI integration, while others rely on pilots and incremental adaptation due to internal capability gaps and high development costs.	&	P5, P6, P9,	P20, P22 \\  \hline 

\end{tabular}
\end{table*}

\textbf{Industry and market challenges} (Table \ref{indu challenges}) in SDV ecosystems arise from the restructuring of value chains, intensified global competition, and the uncertain economics of software-driven mobility. The transition from hierarchical manufacturing systems to networked, software-centric ecosystems requires governance across suppliers, development cycles, and new process models comparable to those used in consumer electronics [P8, P9]. Traditional OEMs face competitive asymmetries as agile newcomers and Chinese automakers leverage lower costs, government support, and integrated digital capabilities, while European manufacturers contend with high energy prices, supply chain disruptions, and slow adaptation [P7, P14]. The development of advanced automation remains expensive, with Level 4 systems requiring substantial investment and lengthy break-even periods [P20]. Collaboration inefficiencies further constrain innovation due to misaligned incentives, data ownership conflicts, and weak cross-industry partnerships [P10, P18]. Uncertainty in data monetisation and dependence on complementors amplify strategic risks [P1, P12], while the adoption of open-source software challenges traditional proprietary business models, forcing the industry to rethink value creation in globally competitive, cost-pressured markets [P14, P15].

\begin{table*}[h]
\caption{Industry \& Market Challenges}
\fontsize{9}{10}\selectfont
\label{indu challenges}
\begin{tabular}{|p{2.7cm}| p{13cm}|p{1.3cm}|}

\hline
\textbf{Category} & \textbf{Description} & \textbf{Source} \\ \hline

Value chain transformation	&	The shift from hierarchical to networked value chains requires extensive coordination and restructuring. Long development lead times and dependence on tiered suppliers slow the transition toward user-centric, software-driven vehicles.	&	P8, P9 	\\  \hline 
Competition and market asymmetries	&	Tesla’s software dominance and the rapid growth of Chinese automakers—supported by policy and cost advantages—intensify competition. European manufacturers are facing rising energy costs, inflation, and productivity challenges that strain their profitability.	&	P5, P7, P14	\\  \hline 
High R\&D and automation costs	&	Developing and validating advanced automation (e.g., Level 4 systems) involves high costs, long timelines, and uncertain ROI, which can make large-scale deployment financially unfeasible for many OEMs and suppliers.	&	P3, P6, P20	\\  \hline 
Collaboration inefficiencies	&	Persistent tensions over data ownership, limited industrial integration, and weak cross-border collaboration hinder effective resource use and innovation. Few joint patents and provisional partnerships reflect low intra-industry cooperation.	&	P10, P18	\\  \hline 
Data-driven business model uncertainty	&	Value migration toward digital complementors, as well as the commoditisation of core vehicle hardware, create uncertainty in monetisation. Companies struggle to capture value from software and data services.	&	P1, P2, P12	\\  \hline 
Open-source business model adaptation	&	Open-source collaboration offer cost and innovation benefits but forces established players to reconsider closed, proprietary business models and redefine value creation within competitive global markets.	&	P14, P15	\\  \hline 

\end{tabular}
\end{table*}

\textbf{Regulatory, legal, and ethical challenges} (Table \ref{reg challenges}) in SDV ecosystems aim to ensure safety, security, and compliance in a fast-evolving and fragmented global landscape. Expanding vehicle connectivity increases the risk of cyberattacks and data breaches, requiring multilayered security frameworks and robust encryption to protect safety-critical systems [P1, P4]. Compliance and certification processes impose strict oversight, extending development timelines and limits flexibility as manufacturers adapt to diverse, evolving standards [P7, P11]. Fragmented international regulation and lack of legal frameworks for Level 4–5 automation hinder large-scale deployment, while inconsistent national oversight limits cross-border certification and enforcement [P14, P19]. Meanwhile, ensuring software reliability and ethical AI behaviour remains unresolved [P5, P7]. Failures in system integrity or driver handover raise safety concerns and regulatory scrutiny [P6, P20], reflecting the broader challenge of governing autonomous decision-making in vehicles that increasingly act without human control.

\begin{table*}[h]
\caption{Regulatory, Legal \& Ethical  Challenges}
\fontsize{9}{10}\selectfont
\label{reg challenges}
\begin{tabular}{|p{2.5cm}| p{13,3cm}|p{1.2cm}|}

\hline
\textbf{Category} & \textbf{Description} & \textbf{Source} \\ \hline

Cybersecurity and data protection	&	SDVs face heightened cybersecurity risks due to extensive connectivity and standardised data exchange. Unauthorised access, hacking, and data breaches threaten safety and privacy, requiring multilayer security and advanced encryption frameworks.	&	P1, P4	\\  \hline 
Compliance and certification burdens	&	Mandatory regulatory compliance, including audited safety and cybersecurity standards, prolong development cycles and limit flexibility. Vehicles failing to meet certification criteria cannot enter the market.	&	P7, P11	\\  \hline 
Fragmented and lagging international regulation	&	Regulatory frameworks for SDVs remain inconsistent and underdeveloped across regions, creating interoperability challenges and slowing deployment of higher automation levels. The lack of harmonised standards for Level 4–5 autonomous systems, and divergent safety, cybersecurity, and environmental requirements, complicates compliance and hinders global market entry.	&	P14, P19, P20	\\  \hline 
Safety, reliability, and ethical implications of AI	&	Ensuring reliable, defect-free software and secure hardware–software integration is critical for public safety. At the same time, AI-driven automation raises unresolved issues of accountability, fairness, and social impact, with regulatory and technical barriers hindering broader societal benefits.	&	P5, P6, P7, P20	\\  \hline 

\end{tabular}
\end{table*}

\subsection{Opportunities (RQ5)}

This systematic review identified not only challenges but also significant opportunities across four key areas, including (1) software, (2) organisational, (3) industry and market, as well as (4) public value and ethical opportunities.

\textbf{Software opportunities} (Table \ref{Software opportunities}) in SDV ecosystems stem from the standardisation, modular architectures, and continuous innovation. Unified data models, standardised APIs, and interoperable platforms simplify system integration and scalability across diverse vehicle models [P1, P22, P25]. Cloud-based and service-oriented architectures enable dynamic resource management and new digital revenue streams [P2, P24], while modular and virtualised systems, supported by continuous OTA updates and digital twins, facilitate rapid feature deployment and lifecycle improvements [P7, P11]. Open-source and SDK-based environments enhance flexibility and developer productivity through reusable components and shared frameworks [P16]. Finally, AI-driven, data-centric architectures supported by scalable E/E systems and edge computing allow continuous learning, performance optimisation, and can reinforce regional technological sovereignty in software and chip development [P14, P19].

\begin{table*}[h]
\caption{Software development Opportunities}
\fontsize{9}{10}\selectfont
\label{Software opportunities}
\begin{tabular}{|p{3cm}| p{12.5cm}|p{1.5cm}|}

\hline
\textbf{Category} & \textbf{Description} & \textbf{Source} \\ \hline

Standardisation, interoperability, and platform integration	&	Establishing standardised data formats, APIs, and software architectures enhances interoperability, simplifies system integration, and supports scalable, modular SDV development across vehicle models.	&	P1, P21, P22, P25 \\  \hline 
Cloud-based and service-oriented architectures	&	Cloud computing and SOAs enable scalable data management, dynamic resource sharing, and new digital business models and revenue streams.	& P2, P24 	\\  \hline 
Modularity, virtualisation, and continuous updates	&	Modular and virtualised architectures, combined with OTA updates, allow flexible deployment, lower lifecycle costs, and faster post-sale feature delivery through continuous integration and improvement.	& P3, P7, P11	\\  \hline 
Open-source and scalable development ecosystems	&	Open, cloud-native, and SDK-based environments empower developers, enhance flexibility, and foster innovation through shared tools, reusable code, and standardised interfaces.	& P16, P24, P25	\\  \hline 
AI-driven, data-centric evolution	&	Continuous learning through networked vehicle fleets, edge computing, and AI integration enables self-optimising systems, performance improvements, and technological sovereignty.	& P14, P19 	\\  \hline 

\end{tabular}
\end{table*}

\textbf{Organisational opportunities} (Table \ref{Org opportunities}) in SDV ecosystems arise from the shift toward software-centric, data-driven, and collaborative operating models. Cloud-based development and SOAs enable real-time updates, operational efficiency, and organisational agility [P2, P3], while data-driven models can open new revenue streams through predictive services and continuous product enhancement [P7, P21]. Modular design, standardisation, and adaptive division of labour can strengthen collaboration across value chains and improve flexibility in manufacturing and sourcing [P2, P10]. OEMs also can balance internal capability building with strategic partnerships, using selective make–buy–ally strategies to retain control over core technologies while accessing external innovation [P12, P13, P15]. Open-source collaboration can further enhance organisational learning, reducing dependency on proprietary systems and accelerating shared innovation [P15]. Platform-based business models allow integration of diverse software, fostering scalable co-development and long-term innovation across the ecosystem [P19, P23].

\begin{table*}[h]
\caption{Organisational Opportunities}
\fontsize{9}{10}\selectfont
\label{Org opportunities}
\begin{tabular}{|p{3cm}| p{12.5cm}|p{1.5cm}|}

\hline
\textbf{Category} & \textbf{Description} & \textbf{Source} \\ \hline

Digital transformation and operational efficiency	&	Transitioning to software-centric and cloud-based operations can improve efficiency, enable real-time updates, and support organisational agility and digital resource management.	& P2, P3, P13	\\  \hline 
Data-driven organisational models	&	Leveraging connected vehicle data can enable predictive maintenance, usage-based insurance, and continuous updates, creating opportunities for new revenue streams and enhanced customer experience.	& P7, P21	\\  \hline 
Value-chain collaboration	&	Modular design, standardized production, and adaptive division of labor can strengthen collaboration between OEMs and suppliers and enhance flexibility across the value chain.	& P2, P9, P10, P20	\\  \hline 
Strategic internalisation and capability development	&	OEMs can balance internal capability building with external collaboration, using selective make–buy–ally strategies and talent-focused approaches to strengthen core competencies.	& P12, P13, P15	\\  \hline 
Open-source collaboration and organisational learning	&	Open-source participation can promote knowledge sharing, innovation, and reduced dependency on proprietary systems while strengthening in-house expertise.	&	P15, P16, P25 \\  \hline 
Platform-based collaboration and innovation	&	Platform-based business and development models integrate software and components from multiple suppliers, which can enable flexibility, innovation, and scalable co-development.	& P19, P22, P23, P24	\\  \hline 

\end{tabular}
\end{table*}

\textbf{Industry and market opportunities} (Table \ref{Ind opportunities}) in SDV ecosystems reflect the transformation of the automotive sector toward software-driven value creation and competitive renewal. Software-based revenue models can enable continuous monetisation through subscriptions, data-driven services, and post-sale features, strengthening customer engagement [P2, P3]. At the same time, open collaboration across OEMs, suppliers, and technology providers can attract new entrants, accelerate shared innovation, and improve efficiency through common digital platforms [P10, P16]. Industry-wide modernisation, driven by automation, digitalisation, and regional initiatives, can enhance manufacturing agility, technological sovereignty, and global competitiveness [P6, P8, P12]. Continuous updates and personalised customer experiences can increase customer satisfaction and brand loyalty, turning vehicles into evolving digital products that maintain long-term market growth [P23].

\begin{table*}[h]
\caption{Industry and Market Opportunities}
\fontsize{9}{10}\selectfont
\label{Ind opportunities}
\begin{tabular}{|p{3cm}| p{12.5cm}|p{1.5cm}|}

\hline
\textbf{Category} & \textbf{Description} & \textbf{Source} \\ \hline

Software-based revenue and digital services	&	SDVs can enable new, recurring revenue streams through subscriptions, data monetisation, and post-sale digital features, strengthening customer engagement and competitiveness.	& P2, P3, P7, P17, P24, P25	\\  \hline 
Ecosystem collaboration and open innovation	&	Open-source development, and cross-industry collaboration can attract new entrants, foster joint innovation, shared resource efficiency, and accelerated digital service creation.	&	P10, P15, P16, P18 \\  \hline 
Industrial modernisation and global competitiveness	&	SDV and automation innovation can improve safety, drive industrial renewal, strengthen regional technological sovereignty, and enhance global competitiveness through new manufacturing and R\&D opportunities.	& P6, P8, P12, P14, P20, P21	\\  \hline 
Customer experience and added value	&	Continuous updates, personalised user experiences, and smart connectivity can enhance brand loyalty, drive customer satisfaction, and expand total market value.	& P3, P5, P17, P23	\\  \hline 

\end{tabular}
\end{table*}

\textbf{Public value and ethical opportunities} (Table \ref{public and ethical opportunities} in SDV ecosystems focus on enhancing safety, sustainability, and societal well-being. Automation, AI, and rigorous testing can reduce human error, improve vehicle reliability, and foster public trust through transparency and accountability [P5, P11]. In turn, connected and automated mobility can advance environmental and social goals by supporting decarbonisation, accessibility, and equitable transport [P6, P21]. These developments not only improve efficiency and reduce congestion but also contribute to broader societal resilience and long-term well-being [P24].

\begin{table*}[h]
\caption{Public Value and Ethical Opportunities}
\fontsize{9}{10}\selectfont
\label{public and ethical opportunities}
\begin{tabular}{|p{3cm}| p{12.5cm}|p{1.5cm}|}

\hline
\textbf{Category} & \textbf{Description} & \textbf{Source} \\ \hline

Safety, reliability, and societal trust	&	Automation, AI, and comprehensive testing can improve road safety, reduce human error, and strengthen public trust through transparent and reliable systems.	& P3, P5, P6, P11, P20, P22	\\  \hline 
Sustainability and societal well-being	&	SDV and automation technologies, along with connected mobility, can support sustainable, accessible, and equitable transportation. These advancements facilitate decarbonization, contribute to broader SDG goals, reduce congestion, and enhance overall societal well-being.	&	P6, P14, P15, P20, P21, P24 \\  \hline 

\end{tabular}
\end{table*}

\section{Discussion}

\subsection{Fragmented Yet Technically Driven SDV Ecosystem Research (RQ1)}

The existing knowledge about SDV ecosystems reveals a technical focus centred around the standardisation of SDV architecture \cite{becker2022operating}, interoperability \cite{Narla2025FrameworkShells}, and cybersecurity \cite{bhandari2024software, rao2023accelerating}. Although the importance of new business modes \cite{unseldshaping, aust2022vehicle, robinson2024ai} and collaborative efforts \cite{Paulweber2025Multi-PartnerEcosystem, merlitz2025open} are recognised, there is a notable lack of empirical studies on governance structures, especially in terms of collaboration around software, business models, and long-term collective prosperity among ecosystem actors. Additionally, there is limited focus on the unique characteristics of SDV ecosystems in different geographic regions. As the industry undergoes transformative changes toward SDVs in various geographical areas, the power dynamics both within and between these ecosystems in different areas are being redistributed (see Table \ref{power}). Without a multi-level perspective and a focus on governance at various levels of these ecosystems \cite{Hietala2025GoverningEcosystems}, this transformation may lead to adverse outcomes. Therefore, in addition to the technical focus, it is equally essential to explore non-technical elements of the ecosystems, such as governance related to software collaboration and the multi-level perspective, which is currently underdeveloped in the literature.

\subsection{Multi-Level Collaborative Structures in SDV Ecosystems (RQ2 and RQ3 + RQ4 and RQ5)}

Our synthesis of the primary articles identified six levels of collaborative structures, including (1) intra-organisational, (2) bilateral/strategic partnerships, (3) supplier and value chain collaboration, (4) open-source collaboration, (5) industry consortia and standardisation networks, as well as (6) policy, research, and cross-sectoral collaboration (Table \ref{structures}). Unlike linear supply chains \cite{xu2020digital}, these collaborative structures connect numerous stakeholders (Table \ref{stakeholders}) across various collaborative constellations, demonstrating the interdependence of these levels. 

This interdependency and need for collaboration around SDV development pose challenges for both the firms involved (Table \ref{org challenges}) and the broader industry (Table \ref{indu challenges}). OEMs encounter risks and inefficiencies when they attempt to develop software in-house without adequate expertise \cite{szalavetz2024inhouse}. On the other hand, outsourcing and relying on open-source solutions can limit their autonomy \cite{szalavetz2024inhouse, becker2022operating, merlitz2025open}. Striking the right balance between proprietary development and collaboration within the ecosystem remains unresolved. Furthermore, transforming the complex value chains necessitates ongoing coordination among OEMs, suppliers, and partners, as well as coordination across multiple disciplines, for aligning hardware, software, and AI development, but existing legacy structures can hinder this co-evolution \cite{bhandari2024software, panchal2023design, lu2023vehicle, lu2024vehicle}. Moreover, challenges such as a lack of expertise, resource constraints, and inconsistent cooperation often impede co-creation and innovation. Further complicating matters are tensions regarding data ownership \cite{liu2022impact}, limited integration, and weak cross-border cooperation \cite{tang2025research}. The migration of value towards digital complementors creates uncertainties around monetisation \cite{szalavetz2024inhouse}. While open-source collaboration can offer cost benefits, it requires established players to rethink proprietary models and value creation in competitive markets \cite{Paulweber2025Multi-PartnerEcosystem, merlitz2025open}. 

Regardless of these challenges, various organisational (Table \ref{Org opportunities}) and industry opportunities (Table \ref{Ind opportunities}) for collaboration between OEMs and suppliers exist through modular design, standardised production, and adaptive labour division \cite{liu2022impact}. OEMs can enhance core competencies by balancing internal capability building with strategic external collaborations, using selective make–buy–ally strategies \cite{szalavetz2024inhouse, laaschmaximize, merlitz2025open}. Embracing open-source participation can foster knowledge sharing, innovation, and reduces reliance on proprietary systems, while strengthening in-house software expertise \cite{merlitz2025open, becker2022operating, wu2024vpi}. Additionally, platform-based business models that integrate software and components from various suppliers can promote flexibility, innovation, and scalable co-development \cite{unseldshaping, unseld2025significant, lu2024vehicle, lu2023vehicle}. In general, open-source development and cross-industry collaboration can attract new entrants, foster joint innovation, enhance resource efficiency, and accelerate digital service creation \cite{liu2022impact, tang2025research}. Innovations in SDV and automation can improve safety, drive industrial renewal, strengthen regional technological sovereignty, and boost global competitiveness through new manufacturing and R\&D opportunities \cite{Paulweber2025Multi-PartnerEcosystem, bhandari2024software}.

However, these collaborative structures are not neutral coordination mechanisms. They are sites of power negotiation that shape who controls architectures, standards, data, and value capture. OEMs have traditionally dominated through system integration and hardware control, but the shift toward SDVs redistributes power toward software suppliers, platform providers, and digital complementors (Table \ref{power}). Control over software architectures, APIs, OSs, and services increasingly determines who holds power within the ecosystem, limiting some actors while providing opportunities for others. Participation in open-source projects and industry consortia can reduce dependency on single vendors \cite{prasanna2023system, merlitz2025open}, but it also necessitates new forms of governance and collective decision-making \cite{Paulweber2025Multi-PartnerEcosystem}, which could also lead to the emergence of new power dynamics, favouring those with greater technical expertise or the ability to influence agendas. As a result, collaboration simultaneously mitigates and reproduces power asymmetries across ecosystem levels.

These shifting power relations directly affect technical and organisational outcomes. Decisions about modularisation, interface openness, and standard adoption become strategic choices through which actors can stabilise or renegotiate their positions within the ecosystem \cite{szalavetz2024inhouse, prasanna2023system}. As a result, organisational structures, capability development, and collaboration modes co-evolve with software architectures. Collaboration thus moves beyond coordination, actively shaping both system design and ecosystem governance. This shift highlights software as the central structuring principle of SDV ecosystems, which is examined in the next section.

\subsection{Software and its development as the Core Structuring Principle of SDV Ecosystems (RQ4 and RQ5)}

As the automotive industry transitions toward SDVs, software is increasingly becoming the primary structuring element of SDV ecosystems, rather than merely a supporting component of vehicle functionality. Software presents opportunities to redesign vehicles around standardised, modular, and SoAs that enable continuous updates, scalable integration, and secure operation across organisational and ecosystem boundaries (Table \ref{Software opportunities}), thereby influencing not only performance and user experience but also collaboration models and control over value creation. At the same time, software-centric architectures introduce new tensions related to how systems are standardised, how software and hardware are modularised and coordinated, how open-source components are integrated into proprietary environments, and how security, integrity, and accountability are ensured in highly interconnected systems (Table \ref{software challenges}). Together, these dynamics position software as the core organising principle of SDV ecosystems, warranting a closer examination of the software considerations identified in this SLR.

\textit{Standardisation and interoperability} are essential for the effective integration in SDV ecosystems. While SDVs rely on common standards, such as AAS, AUTOSAR, and standardised vehicle APIs, to achieve interoperability across various systems, suppliers, and ECUs (cf. \cite{Narla2025FrameworkShells, rao2023accelerating, panchal2023design, aust2022vehicle}), fragmented toolchains and proprietary architectures hinder seamless integration \cite{khamis2025rethinking}. Progress in open standards and unified software–hardware abstraction layers can significantly reduce system fragmentation and integration costs, contributing to the effective development of SDVs within the ecosystem.

\textit{Modular architecture and coordination across software and hardware} are essential for the effective development of SDVs. The shift from hardware-centred to software-defined vehicles demands modular, scalable architectures that enable continuous updates, virtualisation, and reuse across vehicle models \cite{prasanna2023system, panchal2023design}. However, the misalignment of software and hardware development cycles, as well as disciplinary silos, creates delays and inefficiencies in SDV design and development. The decoupling of software from hardware, along with virtualisation, digital twins, and standardised APIs, supports agile and iterative development, faster feature rollout, and consistent system performance across hardware generations \cite{szalavetz2024inhouse, unseld2025significant, prasanna2023system}. This interconnectedness underscores the importance of cohesive software and hardware development in the broader context of SDV ecosystems.

\textit{OSS and ecosystem collaboration} drive innovation and efficiency in the software development value chain. Prior research \cite{SCHREIBER2026112765} shows that OSS collaboration over time is increasingly structured and strategic, with competing firms forming persistent, multiplex, and relatively balanced collaborative ties within the ecosystem, particularly during periods of technological change or standardisation. These so-called coopetitive relationships position OSS as a neutral, pre-competitive coordination mechanism that enables shared innovation without centralised control \cite{SCHREIBER2026112765}. OSS can offer cost efficiency, innovation speed, and reduced vendor dependency, but challenges traditional proprietary industry structures \cite{Paulweber2025Multi-PartnerEcosystem, merlitz2025open}. Sustainable collaboration requires effective governance, vulnerability management, and a balanced approach to the use of open and closed components. When managed properly, OSS enables co-development, modular reuse, and shared innovation across the SDV value chain, thereby enhancing the flexibility and responsiveness of SDV ecosystems to market changes and technological advancements.

\textit{Cybersecurity,  software integrity, and accountability} are critical to safeguard the increasingly connected environments of SDVs. The growing connectivity of SDVs expands the attack surface, through wireless, OBD, and cloud interfaces, requiring multilayered security with encryption, authentication, and embedded firewalls \cite{rao2023accelerating, rehberg2024industrial, bhandari2024software}. Ensuring software integrity demands continuous vulnerability scanning, coordinated countermeasures, and security-by-design principles that maintain trust and resilience in complex digital ecosystems. Yet, as SDVs increasingly rely on open-source and third-party software, accountability becomes a central concern. When defects or security failures cause harm, determining responsibility among OEMs, suppliers, and software contributors remains ambiguous. Addressing this gap requires clear liability frameworks and governance mechanisms to ensure transparency and public trust in safety-critical software systems.

These shifts in software development from an internal engineering concern to an ecosystem-wide coordination challenge, as SDV development relies on shared interfaces, distributed development, and continuous development throughout the vehicle lifecycle. Thus, software is not just a technical layer but the primary organising logic of the ecosystem, calling for novel business and value co-creation models.

\subsection{Business and Value Creation Models in SDV Ecosystems (RQ3 + RQ4 and RQ5)}

\begin{figure*}[h!]
    \centering
    \includegraphics[width=1\linewidth]{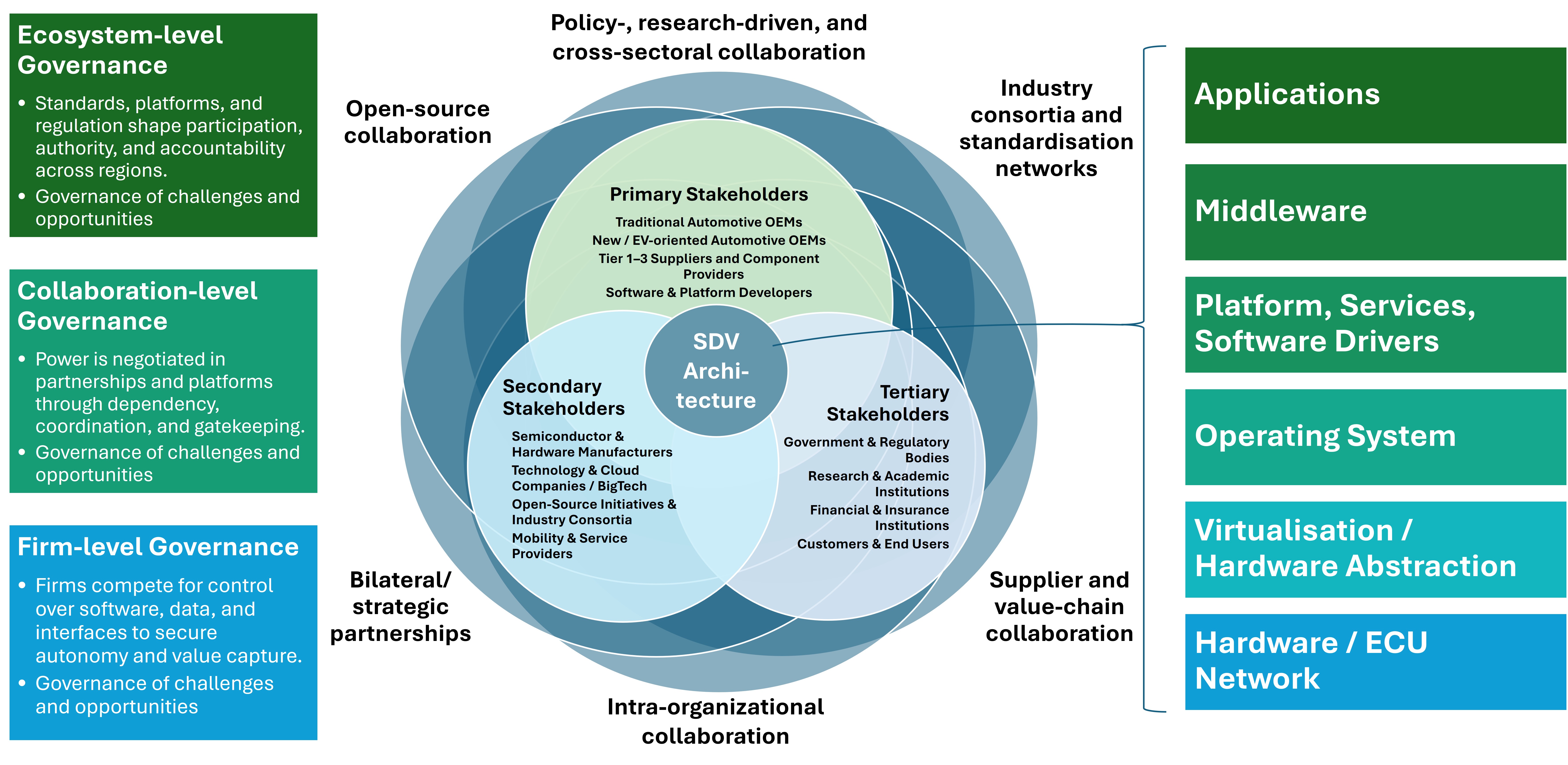}
    \caption{SDV Ecosystem Model}
    \label{SDV ecosystem model}
\end{figure*}

Across SDV ecosystems, business value creation increasingly shifts from vehicle sales toward software-driven services, data monetisation, and platform-based offerings. Organisational opportunities such as software-centric operations, cloud integration, and data-driven models (Table \ref{Org opportunities}) enable OEMs to improve operational efficiency while opening recurring revenue streams through subscriptions, predictive maintenance, and post-sale digital features \cite{liu2022impact, Panchal2023DesignOpportunities, bhandari2024software}. At the industry level (Table \ref{Ind opportunities}), these mechanisms support new market logics in which customer engagement, continuous updates, and ecosystem participation become central to competitiveness. 

The shift toward SDVs is reconfiguring the automotive value chains by enabling direct OEM partnerships that bypass traditional supplier tiers and elevate software capabilities as a core source of competitive advantage. System integrators increasingly act as platform orchestrators, providing transformation services and shared development models across OEMs \cite{BupalamPrasanna2023SystemVehicles}. Similarly, semiconductor firms are moving from component supply to solution provision, as seen in Volkswagen’s CARIAD collaborating directly with chip manufacturers \cite{liebetrau2022architecture}. These changes force OEMs to make strategic make–buy–ally decisions in software development based on differentiation potential and internal capabilities \cite{Liu2022ImpactVehicles, khamis2025rethinking}, while requiring investments in software talent, cultural change, and ecosystem partnerships to remain competitive in software-defined mobility \cite{Liu2022ImpactVehicles}.

Business model realisation in SDV ecosystems increasingly relies on platform-based strategies, phased deployment, and ecosystem orchestration that balance openness with competitive differentiation. Successful approaches emphasise platform-first architectures that enable differentiation through software applications, hybrid open-source models combining open-core platforms with proprietary extensions, and consortium-based development via organisations such as AUTOSAR and COVESA to share pre-competitive investments~\cite{BupalamPrasanna2023SystemVehicles, Rao2023AcceleratingIndustry, merlitz2025open}. Transformation risks are mitigated through pilot-driven migration paths and centralised DevOps project offices that coordinate multi-organisational software development across partner ecosystems~\cite{Rao2023AcceleratingIndustry}. Competitive advantage is protected through intellectual property frameworks and privacy-preserving collaboration mechanisms, such as data desensitisation~\cite{Liu2022ImpactVehicles}. SDV capability development typically combines internal software competency centres, strategic partnerships with technology leaders, and the adoption of open interoperability standards~\cite{BupalamPrasanna2023SystemVehicles}. In this context, system integrators emerge as ecosystem orchestrators, capturing value through shared platform services rather than proprietary control, reflecting a broader shift toward ecosystem-based value creation and capture~\cite{BupalamPrasanna2023SystemVehicles}.

However, value capture remains contested. High R\&D and automation costs, uncertain returns from data-driven business models, and value migration toward digital complementors constrain profitability (Table \ref{indu challenges}). Strategic internalisation, selective open-source collaboration, and modular value-chain reconfiguration assist for retaining control over core assets while leveraging ecosystem innovation. Public value creation, through improved safety, trust, and sustainability, further shapes business viability (Table \ref{public and ethical opportunities}), as regulatory compliance, societal acceptance, and ethical accountability increasingly condition market access and long-term ecosystem prosperity (Table \ref{reg challenges}).

Authority and power distribution in SDV ecosystems are closely tied to platform control, as actors that govern software platforms increasingly determine value capture. OEMs face persistent tensions between internalising software capabilities and relying on suppliers, while asymmetric dependencies raise the risk of value migration toward Big Tech and cloud providers. These dynamics are reinforced by high R\&D costs, collaboration inefficiencies, and market asymmetries, even as new revenue streams, ecosystem-level innovation, and enhanced customer value emerge. Under conditions of uncertainty, path dependencies, and strategic lock-ins, early platform and governance choices can have long-lasting implications for competitive positioning and ecosystem viability.

\subsection{Theoretical Contribution: A Multi-Level SDV Ecosystem Model}

This study advances SDV ecosystem research by challenging the prevailing view of SDVs as primarily a technical standardisation problem. While the existing literature provides rich insights into open-source development, interoperability standards, and architectural frameworks, it has not yet explored how these technical layers are embedded in broader organisational, collaborative, and governance arrangements. In response, we reposition SDVs as multi-level socio-technical ecosystems with technical realisation serving as a foundation for the ecosystems' success.

Drawing on parallels from digital twin ecosystems \cite{tripathi2024stakeholders, xu2020digital} and digital public service ecosystems \cite{Hietala2025GoverningEcosystems}, we develop a conceptual model that integrates key stakeholders, collaborative constellations, and governance requirements, including power dynamics, across firm, collaboration, and ecosystem levels (Figure~\ref{SDV ecosystem model}). This model emphasises the importance of recognising and governing power dynamics, control mechanisms, and decision-making processes, as described in Section 4.3.3, alongside technical interoperability. Thus, we argue that the effectiveness of ecosystem development depends not only on aligned incentives and governance mechanisms but also on acknowledging the complexities of power relations among key actors. 

By addressing these dynamics, we identify both challenges and opportunities within the ecosystem. For instance, where power imbalances exist, collaboration may become strained, and value creation could be jeopardised. Conversely, by fostering inclusive decision-making and equitable control structures, ecosystems can leverage diverse perspectives, enhancing innovation and long-term prosperity. Thus, this study clarifies how governance alignment, informed by an understanding of power and decision-making, is crucial for shaping innovation, value creation, and sustainable development in SDV ecosystems.

\subsection{Implications to Research}

This SLR identifies several directions for future SDV ecosystem research. First, SDV ecosystems differ fundamentally from mobile phone ecosystems due to substantially higher safety, cybersecurity, and regulatory requirements, as well as tighter interdependencies between hardware, software, and operational contexts. Software failures in SDVs can have direct physical and societal consequences, which highlights the importance of studying interdependencies and governance mechanisms that span technical, organisational, and regulatory domains.

Second, SDV ecosystems are shaped by regional regulatory regimes and policy incentives. While existing research largely focuses on standardisation and technological development, there is limited insight into how SDV ecosystems emerge and evolve under different regulatory conditions. To gain a clearer picture of how unique geographical and regulatory contexts affect the structures of these ecosystems, collaboration patterns, and innovation trajectories, further research across diverse regions is necessary.

Third, a strong focus on technical building blocks may accelerate SDV development, yet long-term ecosystem success depends equally on governance and collaboration. Historical cases like Nokia, Ericsson, and BlackBerry demonstrate that technological advancement alone does not guarantee long-standing competitiveness (cf. \cite{hari2014impact}). This is particularly salient for incumbent OEMs, which must adapt their roles, partnerships, and governance arrangements to avoid becoming obsolete. Future research should examine how multilevel governance and coordination can be structured to support durable SDV ecosystems.

Fourth, OSS is widely viewed as central to SDV development, yet it raises unresolved challenges related to safety, reliability, and accountability. Asymmetric participation, where OEMs primarily use rather than contribute to open-source solutions, may hinder collective progress \cite{merlitz2025open}. Future research is needed on governance models that ensure safety and accountability in open collaboration, incentivise active OEM participation, and support sustainable transitions from proprietary to open-source SDV architectures.

Finally, SDV ecosystems are not only networks of firms and technologies but also socio-technical systems that directly affect human safety, trust, and fairness. Although the human dimension is frequently mentioned in relation to safety, user experience, and human–machine interaction, there are significant research opportunities to explore it more deeply. While concepts such as human-centred autonomous systems exist \cite{panchal2023design}, more ecosystem-level research is needed to understand how human considerations are embedded in SDV design, governance, and collaboration. Addressing this gap is essential for developing SDV ecosystems that support safe, inclusive, and sustainable mobility over the long term.

\subsection{Implications to Practice}

Organisations should pay close attention to aligning governance across firm, collaboration, and ecosystem levels, as fragmented decision-making can undermine SDV development. Strategic choices regarding internalisation versus partnerships are critical, particularly in balancing control over software and data with access to external capabilities. Active participation in open-source and standardisation networks requires active contribution, as limited commitment by key actors can undermine collective progress and threaten ecosystem viability. Clear accountability frameworks must be established across software supply chains to address responsibility for failures, security breaches, and compliance. In addition, organisations need inter-organisational coordination mechanisms that support collaboration beyond technical architectures alone. Without these measures, SDV initiatives risk fragmentation, strategic lock-in, limited ecosystem participation, and weak collective cybersecurity practices.

\section{Conclusion}
\label{sec:conclusion}

As the automotive industry transitions to SDVs, there is a growing need for a deeper understanding of the ecosystems that emerge from this change. This SLR reveals that SDV ecosystems depend on various interconnected collaboration structures, ranging from internal software units within OEMs to bilateral partnerships with cloud and AI firms, supplier networks, open-source communities, standardisation consortia, and policy-driven coordination initiatives. The effectiveness of these collaborations is influenced by five dimensions of authority, highlighting how power, control, and decision-making are being redistributed as the industry evolves. However, SDV ecosystems also face challenges, including software-related, organisational, market and industry, as well as regulatory, legal, and ethical issues. Despite these challenges, the transformation presents significant opportunities in software, organisational, industry and market, and public value and ethical dimensions. By synthesising these insights, this study repositions SDVs as multi-level socio-technical ecosystems in which software functions as the core structuring principle but does not alone determine ecosystem success. In response, the study proposes an SDV ecosystem model centred on software that integrates stakeholders, collaborative structures, governance, and power dynamics across ecosystem levels. These findings show that SDV ecosystems are shaped not only by technological factors but also by the interplay among technical foundations, multi-level governance, coordination, and collaboration.

\section*{Acknowledgment}

The work has been supported by the EU HORIZON projects CHIPS-JU CIA FEDERATE (grant number 101139749), CHIPS-JU RIA HAL4SDV (grant number 101139789) and CHIPS-JU IA SHIFT2SDV (grant number 101194245), Business Finland projects 6G Visible (grant number 10743/31/2022), HAL4SDV national funding (grant number 7655/31/2023) and SHIFT2SDV national funding (5972/31/2024).

\section*{Declaration of generative AI and AI-assisted technologies in the manuscript preparation process}

The authors used ChatGPT 5.2 and Grammarly to copy-edit the text. After using these tools, the authors reviewed and edited the content as needed and take full responsibility for the content of the published article.
\bibliographystyle{abbrv}

\bibliography{references.bib}







\end{document}